\documentclass{aa}
\input{psfig}

\def\tvi(#1,#2){\vrule height #1pt depth #2pt width 0pt}
\def\p{\partial}
\def\e{{\rm e}}
\def\d{{\rm d}}
\def\ie{i.e. }
\def\eg{e.g. }
\def\etal{et al. }
\def\rsh{{r}_{\rm sh}}
\def\rso{r_{\rm son}}
\def\M{{\cal M}}
\def\omcut{\omega_l^{\rm cut}}

\def\omcutu{\omega_1^{\rm cut}}
\def\rc{{\cal R}^{*}}
\def\rt{r_{\rm t}}
\def\eiwv{\e^{i\omega\int_{\rsh}^r {\d r\over v}}}

\newlength{\largeur}
\newlength{\saut}
\def\marge#1{
\setlength{\largeur}{\columnwidth}
\addtolength{\largeur}{-#1}
\setlength{\saut}{0.5\largeur}\hspace*{\saut}} \def\picture #1 by #2
(#3){
\marge{#1} \vbox to #2{
\hrule width #1 height 0pt depth 0pt
\vfill
\special{picture #3}}}
\def\scaledpicture #1 by #2 (#3 scaled #4){{
\dimen0=#1 \dimen1=#2
\divide\dimen0 by 1000 \multiply\dimen0 by #4 \divide\dimen1 by 1000
\multiply\dimen1 by #4 \picture \dimen0 by \dimen1 (#3 scaled #4)}}

\begin{document}

\thesaurus{06
(02.01.2;
02.08.1;
02.09.1;
02.19.1;
08.02.1;
13.25.5)}

\title{Non-radial instabilities of isothermal Bondi accretion with a 
shock: vortical-acoustic cycle vs post-shock acceleration}
\author{T. Foglizzo\thanks{e-mail: {\tt foglizzo@cea.fr}} 
}


\institute {Service d'Astrophysique, CEA/DSM/DAPNIA, CEA-Saclay, 91191
Gif-sur-Yvette, France 
}

\date{Received 27 May 2002 / Accepted 17 June 2002}
\titlerunning{vortical-acoustic cycle vs post-shock acceleration}
\maketitle

\begin{abstract}
The linear stability of isothermal Bondi accretion with a shock is 
studied analytically in the asymptotic limit of high incident Mach number 
$\M_{1}$. The flow is unstable with respect to radial perturbations as 
expected by Nakayama (1993), due to post-shock acceleration.
Its growth time scales like the advection time from the shock $\rsh$ 
to the sonic point $\rso$. The growth rate of non-radial perturbations 
$l=1$ is higher by a factor $\M_{1}^{2/3}$, and is therefore intermediate 
between the advection and acoustic frequencies.
Besides these instabilities based on post-shock acceleration, our 
study revealed another generic mechanism based on the cycle of acoustic and 
vortical perturbations between the shock and the sonic radius, 
independently of the sign of post-shock acceleration. 
The vortical-acoustic instability is fundamentally non-radial. It
is fed by the efficient excitation of vorticity waves by the isothermal 
shock perturbed by acoustic waves. The growth rate exceeds the advection 
rate by a factor $\log\M_{1}$.
Unstable modes cover a wide range of frequencies from the  
fundamental acoustic frequency $\sim c/\rsh$ up to a cut-off $\sim c/\rso$ 
associated with the sonic radius. The highest growth rate is reached 
for $l=1$ modes near the cut-off. The additional cycle of acoustic waves 
between the shock and the sonic radius is responsible for variations of the 
growth rate by a factor up to $3$ depending on its phase relative to the 
vortical-acoustic cycle. 
The instability also exists, with a similar growth rate, below the fundamental 
acoustic frequency down to the advection frequency, as vorticity waves are 
efficiently coupled to the region of pseudosound.
These results open new perspectives to address the stability of 
shocked accretion flows.


\end{abstract}

\section{Introduction}

Hydrodynamic instabilities in accretion flows can help understanding
the variability observed in the luminosity of X-ray binaries.
Numerical simulations have revealed the existence of such a
hydrodynamical instability in the accretion flow of a gas on a compact
accretor moving at supersonic velocity (Bondi-Hoyle-Lyttleton
accretion). A first step towards understanding
the physical mechanism underlying this instability was made by Foglizzo
\& Tagger (2000, hereafter FT00) who recognized the unstable cycle of 
entropic and acoustic perturbations between the shock and the sonic surface. 
This cycle is unstable if there is a large enough temperature difference
between the shock and the sonic surface. The academic case of shocked
Bondi accretion was studied by Foglizzo (2001, hereafter F01) who revealed the
importance of non radial perturbations, and vorticity in particular.
Both vorticity and entropy perturbations are advected from the shock
to the accretor, and both are coupled to the acoustic perturbations.
This coupling was formulated in a compact way by Howe (1975).
If the adiabatic index is in the range $1<\gamma\le5/3$, entropy and
vorticity perturbations are intimately related in the shocked Bondi
flow (Foglizzo 2002, in preparation), so that it is difficult to identify their
respective roles in the instability mechanism. By contrast, their roles
are well separated in the isothermal limit ($\gamma=1$), where entropy
perturbations are absent from the problem. The present paper is
therefore dedicated to the study of the linear stability of shocked
accretion in the isothermal Bondi flow, where the incident Mach number is the
only parameter. The stability of shocked isothermal flows was studied 
by Nakayama (1992, 1993) in the more general context of flows with 
small angular momentum. The study of Nakayama was restricted to 
axisymmetric perturbations, thus precluding any possible vortical 
acoustic cycle. In this approximation, Nakayama analytically obtained 
the result that the flow is unstable if the flow accelerates immediately
after the shock surface. In this respect the shocked Bondi flow should 
be unstable. We are therefore interested here in an
extension of Nakayama's results to the case of non radial
perturbations, in order to include the effect of vorticity. 
It should be noted that isothermal flows are very particular as far as 
Bondi-Hoyle-Lyttleton (hereafter BHL) accretion is concerned, with unsettled 
issues concerning the influence of the numerical resolution.
The instability observed in numerical simulations seems very weak in 3-D 
according to Ruffert (1996), whereas it is violent in 2-D (Ishii \etal 1993, 
Shima \etal 1998). Pogorelov \etal (2000) discussed the possible responsability
of the numerical procedure in producing the instability. With a 
different approach, the instability of the shocked Bondi flow described in 
this paper could contribute to guide our physical understanding of more 
complex flows involving isothermal shocks.

The paper is organized as follows. Perturbed equations are described in
Sect.~\ref{Seq} and eigenfrequencies are determined numerically in 
Sect.~\ref{sect3}. 
Analytical methods are used to disentangle the effect of advection from the 
effect of the boundary. In the spirit of F01, the 
coupling between the vorticity and acoustic perturbations in the classical 
isothermal Bondi flow, without a shock, is described in 
Sect.~\ref{Svortexsound}. Boundary conditions are 
taken into account to build a vortical acoustic-cycle in 
Sects.~\ref{secvai} and \ref{Sectps}. The instability due to post-shock 
acceleration is analysed in Sect.~\ref{secps}.
The relationship between the vortical-acoustic instability and existing 
numerical simulations of BHL accretion and shocked discs is discussed in 
Sect.~\ref{Sdiscuss}.

\section{Linearized equations of the shocked Bondi flow\label{Seq}}

\subsection{Properties of the unperturbed shocked flow\label{sectprop}}

\begin{figure}
\psfig{file=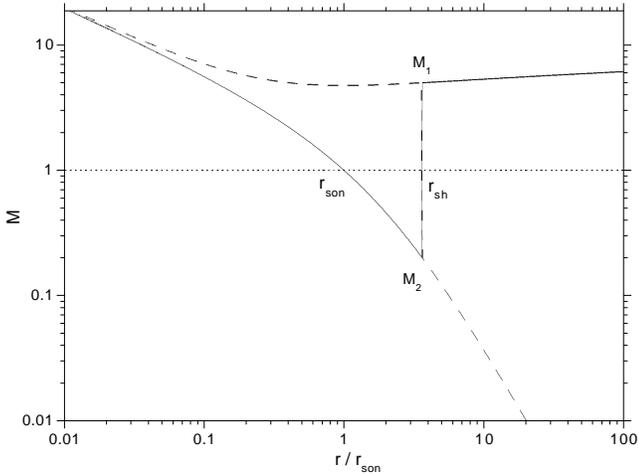,width=\columnwidth}
\caption[]{Radial profile of the Mach number in the unperturbed 
shocked flow, for $M_{1}=5$ (full line). The subsonic cavity stands between 
the sonic point and the shock radius. The  dashed lines show the analytical 
continuation of the solutions beyond the shock radius. }
\label{figsuper}
\end{figure}
The Mach number in the unperturbed transonic flow satisfies the following 
equation deduced from the conservation of the mass flux and Bernoulli 
constant $B$ and the regularity at the sonic radius:
\begin{equation}
{r\over\rso}\M^{1\over2}\exp\left({\rso\over r}-{\M^{2}\over 4}\right)=
\e^{3\over4}.
\label{flow}
\end{equation}
The sonic radius is half of the Bondi radius $GM/c^{2}$. The 
Bernoulli constant,
\begin{equation}
B\equiv {v^{2}\over2} + c^{2}\log\rho -{2\rso\over r},\label{bernoulli}
\end{equation}
is conserved along the flow lines. By contrast with the
adiabatic case $\gamma>1$, this quantity is not conserved through a shock.
The shock radius $\rsh$ corresponding to an incident Mach number 
$\M_{1}$ is determined from Eq.~(\ref{flow}) together 
with the Rankine Hugoniot jump condition $\M_{2}=1/\M_{1}$.
The Mach number profile is shown in Fig.~\ref{figsuper}. It should be 
noted that the presence of a radial shock is not guaranteed a priori, 
since the supersonic preshock flow ($r>\rsh$) could be continued without a 
shock down to the accretor (dashed line in Fig.~\ref{figsuper}). 
Several physical mechanisms could trigger the formation of a shock, such as the 
heating of protons to temperatures at which the fluid becomes collisionless 
(Mezaros \& Ostriker 1983), the trapping of relativistic particules 
(Protheroe \& Kazanas (1983) or the dissipation of magnetic fields 
(McCrea 1956, Scharlemann 1981). 
In the context of Bondi-Hoyle-Lyttleton accretion, the existence of a 
stagnation point behind the accretor implies that a fraction of the supersonic 
flow decelerates to subsonic velocities, and therefore naturally produces 
a shock. The shock radius deduced from Eq.~(\ref{flow}) 
increases with $\M_{1}$ for spherical accretion:
\begin{equation}
{\rsh\over\rso}\sim \e^{3\over4}\M_{1}^{1\over2}.\label{shson}
\end{equation}
Compared to BHL accretion, the presence of an accretion shock at a distance 
exceeding the accretion radius $2GM/v_{\infty}^2$ is rather artificial and 
due simply to the assumption of purely radial velocity. In the highly
supersonic limit $\M_{1}\gg1$ initially considered by Hoyle \& Lyttleton (1939), 
any trajectory with an angular momentum larger than $\sim 2GM/v_{\infty}$ would 
of course miss the accretor. Bearing this feature in mind, the case of strong 
shocks studied in this paper is still very useful in order to understand the 
mechanisms involved, because of the separation of timescales it enables.

\subsection{Scaling of timescales in the shocked Bondi 
flow\label{secscale}}

If the shock is strong, the timescale $\tau_{\rm adv}$ associated with
advection is much longer than the sound crossing time $\rsh/c$:
\begin{equation}
\tau_{\rm adv}\equiv\int^{\rsh}_{\rso} {\d r\over | v|}\sim 
{\M_{1}\over 3}{\rsh\over c}\gg {2\rsh\over c} .\label{tauadv}
\end{equation}
The third fundamental timescale of the flow is related to the presence 
of the sonic radius $\sim\rso/c$. 
Together with Eq.~(\ref{shson}), the scaling of the three special 
timescales of the problem is thus the following for strong shocks:
\begin{equation}
\tau_{\rm adv}\gg{\rsh\over c}\gg {\rso\over c}.\label{scale}
\end{equation}

\subsection{Differential system for the perturbations of the Bondi flow}

\begin{figure}
\psfig{file=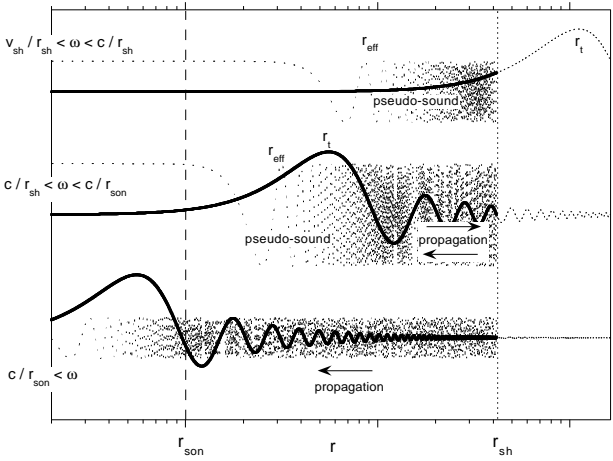,width=\columnwidth}
\caption[]{Schematic structure of the perturbed acoustic (full line) and 
vorticity (dotted line) fields, depending on the frequency of the perturbation. 
The curve beyond the shock radius shows the acoustic perturbation in the Bondi 
flow without a shock. The turning point $\rt$ separates the 
region of pseudosound from the region of propagation of acoustic 
pertrbations. The vortical-acoustic coupling is most efficient in the 
region between $\rso$ and $r_{\rm eff}$.}
\label{figpseudo}
\end{figure}
As in the case $\gamma>1$ studied by F01, the 
vorticity equation can be directly integrated, so that the three quantities 
($r^{2}\delta w_{r},rv\delta w_{\theta},rv\delta w_{\varphi}$) are 
conserved when advected.
The perturbations of the radial velocity $\delta v_{r}$ and density
$\delta \rho$ in the isothermal Bondi flow are conveniently described by
the functions $f,g$ defined by:
\begin{eqnarray}
f&\equiv& v\delta v_{r}+c^{2}{\delta\rho\over\rho},\label{deff}\\
g&\equiv& {\delta v_{r}\over v}+{\delta\rho\over\rho}.\label{defg}
\end{eqnarray}
They satisfy the same differential equations as the functions $f,g$ in 
the Appendix~B of F01, by suppressing the entropy terms:
\begin{eqnarray}
v{\p f\over\p r}+{i\omega \M^2f\over 1-\M^2}
&=& {i\omega v^2 g\over 1-\M^2},\label{dfdri}\\
v{\p g\over\p r}+{i\omega \M^2g\over 1-\M^2}
&=& {i\omega \mu^{2}f\over c^2(1-\M^2)} +
{i\delta K_{\rm sh}\over r^2\omega }\eiwv,\label{dgdri}
\end{eqnarray}
where the function $\mu(r,\omega,l)$ is defined by:
\begin{eqnarray}
\mu^{2}&\equiv&1 - {\omega_{l}^{2}\over \omega^{2}}\label{defmu},\\
\omega_{l}^2&\equiv&{l(l+1)\over r^2}(c^{2}-v^{2}).\label{defomegal}
\end{eqnarray}
The constant $\delta K$ is related to vorticity as follows
\begin{equation}
\delta K\equiv r^{2}v\cdot(\nabla\times w).\label{defKi}
\end{equation}
By contrast with the case of adiabatic flows (F01), $\delta K$ is the unique 
source term for the excitation of acoustic waves.
For radial perturbations ($l=0, \delta K=0$), the equations
analysed by Nakayama (1992, 1993), with $\Psi\equiv f/i\omega$, are 
recovered.\\
The position of the turning point $\rt$ of non-radial acoustic 
waves is defined by $\mu=0$ in Eq.~(\ref{defmu}). This turning point 
lies inside the subsonic cavity if $\omega>\omega_{\rm sh}$, with
\begin{equation}
\omega_{\rm sh}\equiv l^{1\over2}(l+1)^{1\over2}(1-\M_{\rm sh}^{2})^{1\over 2}
{c\over \rsh}\propto {c\over\rsh}.\label{omsh}
\end{equation}
Below $\omega_{\rm sh}$ (or inside the turning point $\rt$ if it 
exists), the acoustic perturbation is not propagating. It was named 
"pseudosound" by Ffowcs Williams (1969).\\
The threshold between absorbed and trapped sound was introduced in F01 
through the cut-off frequency $\omcut$, which is independent of the 
shock strength:
\begin{equation}
\omcut \sim {l^{1\over2}(l+1)^{1\over2}\over2}{c\over \rso}\propto {c\over\rso}.
\label{omcut}
\end{equation} 
This cut-off frequency used to be 
refered to as a "refraction" cut-off in F01 for $\gamma>1$, due to the 
non-homogeneity of the sound speed which bends trajectories outwards. 
By contrast, in the isothermal flow 
considered here, acoustic trajectories are only bent inwards, due to the 
sole effect of flow acceleration. The existence of a cut-off for 
non-radial acoustic waves can be 
understood by considering the frequency dependence of the direction 
$\psi$ of propagation of the incoming acoustic flux relative to the radial 
direction. This angle is approximated in Appendix~E in the WKB approximation 
(Eq.~\ref{defpsi}). The higher the frequency, the smaller $\psi$, so that 
high frequency waves with a given order $l$ are pointing towards the 
accretor and are absorbed inside the sonic sphere.\\
The scaling of frequencies stressed in Eq.~(\ref{scale}) for strong shocks 
consequently separates three ranges of eigenmodes schematized 
in Fig.~\ref{figpseudo}:
\par(i) absorbed sound $\omcut<\omega$,
\par(ii) trapped sound $\omega_{\rm sh}<\omega<\omcut$,
\par(iii) pseudosound $1/\tau_{\rm adv}<\omega<\omega_{\rm sh}$.
The corresponding structure of the vorticity wave is also displayed in 
Fig.~\ref{figpseudo}, in each range of frequencies.

\subsection{Boundary condition at the shock}

The boundary conditions at the shock associated with the variables
$f,g$ can be computed with the same method as Nakayama (1992), extended to 
the case of non radial perturbations.
Let us denote by $\Delta\zeta$ the radial displacement of the shock produced 
by a sound wave propagating against the stream in the subsonic region , and 
$\Delta v=-i\omega\Delta\zeta$ its velocity. The incident Mach number 
$\M_{1}'$ in the frame of the shock is altered by the perturbations 
of velocity and displacement as folllows:
\begin{eqnarray}
\M_{1}'(\rsh+\Delta\zeta)&=&\M_{1}(\rsh)+{\Delta v\over c}+
\Delta\zeta{\p\M_{1}\over\p r},\label{zetav}\\
&=&\M_{1}(\rsh)+\left(1-{i\eta c\M_{2}\over\omega\rsh}\right)
{\Delta v\over c},
\end{eqnarray}
where the parameter $\eta$ measures the strength of the local 
gradient of the Mach number immediately after the shock:
\begin{equation}
\eta\equiv {\p\log \M_{2}\over\p \log r}=-{2\over 1-\M_{2}^{2}}
\left(1-{\rso\over\rsh}\right),
\label{defzeta}
\end{equation}
The boundary conditions $f_{\rm sh},g_{\rm sh}$, written as functions of 
$\Delta v$, are deduced from the conservation of the mass flux and impulsion 
across the shock:
\begin{eqnarray}
f_{\rm sh}&=&-c^{2}\M_{2}(1-\M_{2}^{2})
\left\lbrack(1-\M_{2}^{2}){i\eta c\over\omega\rsh}+\M_{2}\right\rbrack
{\Delta v\over v_{2}},\label{boundaf}\\
g_{\rm sh}&=&(1-\M_{2}^{2}){\Delta v\over v_{2}}.\label{boundag}
\end{eqnarray}
The non radial perturbation of the velocity, and the perturbed 
vorticity computed in Appendix~A enable us to relate
$\delta K_{\rm sh}$ to $\Delta v$:
\begin{equation}
\delta K_{\rm sh}=l(l+1)c^{2}(1-\M_{2}^{2})^{2}
\left( 1-{i\eta c\M_{2}\over\omega\rsh}\right){\Delta v\over 
v_{2}}.\label{eqKsh}
\end{equation}
Eq.~(\ref{eqKsh}) can be used together with 
Eqs.~(\ref{boundaf}-\ref{boundag}) in order to express 
$f_{\rm sh},g_{\rm sh}$ as functions of $\delta K_{\rm sh}$ for 
non-radial perturbations.
In Appendix~C the boundary value problem is reduced to a single 
equation incorporating the boundary conditions both at the shock and 
at the sonic point. The eigenfrequencies satisfy the following 
equation, where $f_{0}$ is the unique homogeneous solution which is 
regular at the sonic point:
\begin{eqnarray}
{\p^{2} f_{0}\over\p r^{2}}+\left\lbrack
{\p\log\over\p r}\left({1-\M^{2}\over\M}\right)-{2i\omega \M\over 
(1-\M^{2})c}\right\rbrack{\p f_{0}\over\p r}\nonumber\\
+{{\omega^{2}\over c^{2}}-{L^{2}\over r^{2}}\over 1-\M^{2}}f_{0}=0,
\label{homog}\\
\left\lbrack (1-\M_{2}^{2})\eta - 
\M_{2}{i\omega \rsh\over c}\right\rbrack {\p f_{0}\over\p r}(\rsh)
\nonumber\\
+{i\omega \rsh\over c}\left\lbrack {1+\M_{2}^{2}\over 1-\M_{2}^{2}}
{i\omega \rsh\over c}
-\M_{2}\eta\right\rbrack 
{f_{0}(\rsh)\over \rsh}=\nonumber\\
l(l+1) \left({i\omega \rsh\over c}+\M_{2}\eta\right)
\int_{\rso}^{\rsh} {f_{0}\over\M r^{2}}
\e^{i\omega\int_{\rsh}^r{1+\M^{2}\over 1-\M^{2}}{\d r\over v}
}\d r.\label{equnique}
\end{eqnarray}
Eq.~(\ref{equnique}) is independent of the normalization of $f_{0}$. 
It should be noted that the integral on the right hand side is 
well defined despite the singularity of the phase near the sonic radius.
This equation describes the perturbation of the shock by the interplay of the 
acoustic and vortical perturbations. The same calculation in any other 
potential (\eg the Paczynski-Wiita potential) would lead to the same 
system of equations (\ref{homog}-\ref{equnique}), only the shape of $\M$ 
(and its derivative $\eta$) described by Eq.~(\ref{flow}) would be affected. 
The left hand side of Eq.~(\ref{equnique}) involves only the acoustic 
perturbation $f_{0}$, and is independent of the vortical perturbation. 
By contrast, the integral on the right hand side describes the acoustic feed back of the vortical 
perturbation coupled to the acoustic field.

\section{Spectrum of eigenfrequencies \label{sect3}}

\subsection{Numerical procedure to determine the spectrum of eigenfrequencies}

The regularity of the solution at the sonic radius has already been discussed
in F01: the sonic point is a regular singularity if $\gamma<5/3$.
For a given incident Mach number $\M_{1}$, numerical intergration 
is performed from 
the sonic point towards the shock in order to determine the eigenfrequencies 
of perturbations with a latitudinal number $l$. For a
given value of $\omega$, the unique solution which is regular at the
sonic point is expanded in a Frobenius series in order to start the
numerical integration away from the singularity. A Runge-Kutta
algorithm is then used to simultaneously integrate four functions, 
namely $f_{0}(\rsh)$, $g_{0}(\rsh)$, the integral on the right hand side of 
Eq.~(\ref{equnique}) and the integral phase inside it. $\omega$ is
varied and shooting is repeated until Eq.~(\ref{equnique}) is satisfied.

\subsection{Purely growing radial instability}

\begin{figure}
\psfig{file=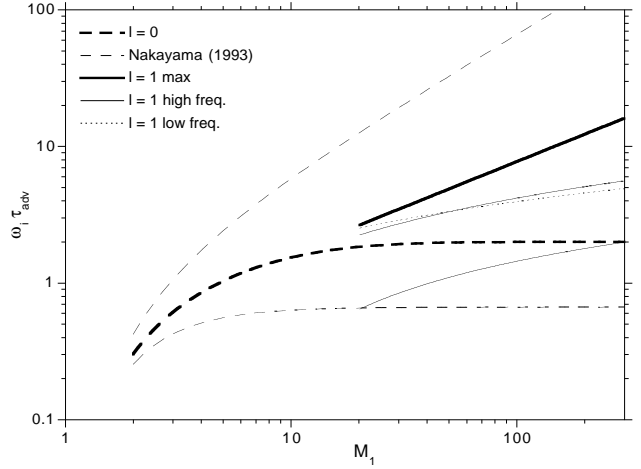,width=\columnwidth}
\caption[]{Growth rate of the purely growing radial instability, 
multiplied by the advection time (thick dashed line). The thin dashed lines 
correspond to the lower and upper bounds determined by Nakayama (1993). 
Anticipating on the calculations of the rest of the paper,
the four other curves are analytical estimates of the growth rate of 
$l=1$ perturbations at high Mach number. The growth rate of the 
entropic-acoustic instability increases along the dotted line at low 
frequency (Eq.~\ref{lowfwi}), and is bounded by the two thin lines
at high frequency (Eqs.~\ref{omm2}-\ref{omp2}). The thick full line describes the 
most unstable branch (Eq.~\ref{maxwi}).  }
\label{fignaka0}
\end{figure}
The radial instability found numerically in Fig.~\ref{fignaka0} is consistent 
with the result of Nakayama (1992, 1993), who found that the local acceleration 
of the flow immediately after the shock is a source of non-oscillating radial 
instability. 
The growth rate is shown as a function of the incident Mach number, 
in units of the advection time $\tau_{\rm adv}$. 
The lower and upper bounds found analytically by Nakayama (1993, Eq.~27)
are also displayed in Fig.~\ref{fignaka0}:
\begin{equation}
{|\nu|\over1+\M_{1}^{2}}\le\omega_{i}\le{|\nu|\over1+\M_{1}},\label{rangenaka}
\end{equation}
where the parameter $\nu$ describes the acceleration of the flow and is 
proportionnal to $\eta$:
\begin{eqnarray}
\nu&\equiv &-{1\over v_{2}}\left({\p\Phi\over\p r}-{2c^{2}\over\rsh}\right),
\label{nuisoth}\\
&=&(\M_{1}^{2}-1){\eta c\M_{2}\over\rsh}.\label{etanu}
\end{eqnarray}
Note that Eq.~(\ref{nuisoth}) is corrected for a factor 2 for spherical flows, 
as noticed by Nakayama (1993). The lower and upper bounds of 
Eq.~(\ref{rangenaka}) cover a wide range of frequencies from the advection 
frequency $\sim 2/(3\tau_{\rm adv})$ up to the acoustic frequency $\sim 2c/\rsh$. 
In the present case of isothermal radial accretion, the growth rate computed 
numerically converges precisely towards the value $2/\tau_{\rm adv}$. 

\subsection{Non-radial oscillatory instabilities}

\begin{figure}
\psfig{file=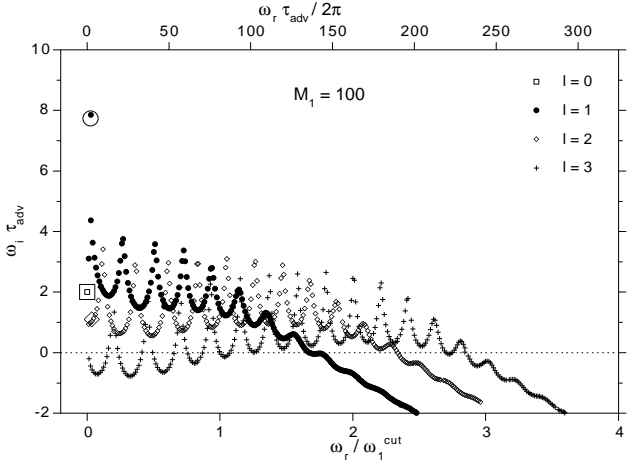,width=\columnwidth}
\caption[]{Real and imaginary parts of the eigenfrequencies of the modes 
$l=0,1,2,3$ in the isothermal flow, for $\M_1=100$ 
($c\tau_{\rm adv}/4\pi\rsh\sim 2.8$). The real part is measured in units of 
the cut-off frequency $\omcutu$ (lower axis) and advection frequency 
(upper axis). The imaginary part is normalized to the advection time. 
Analytical estimates of the growth rates of the modes $l=0,1,2$ at 
low freqency are indicated using big symbols (Eqs.~\ref{wi2}, \ref{sol0} and 
\ref{maxwi} in Sects.~\ref{Sectps} and \ref{secps}). }
\label{figM100}
\end{figure}
\begin{figure}
\psfig{file=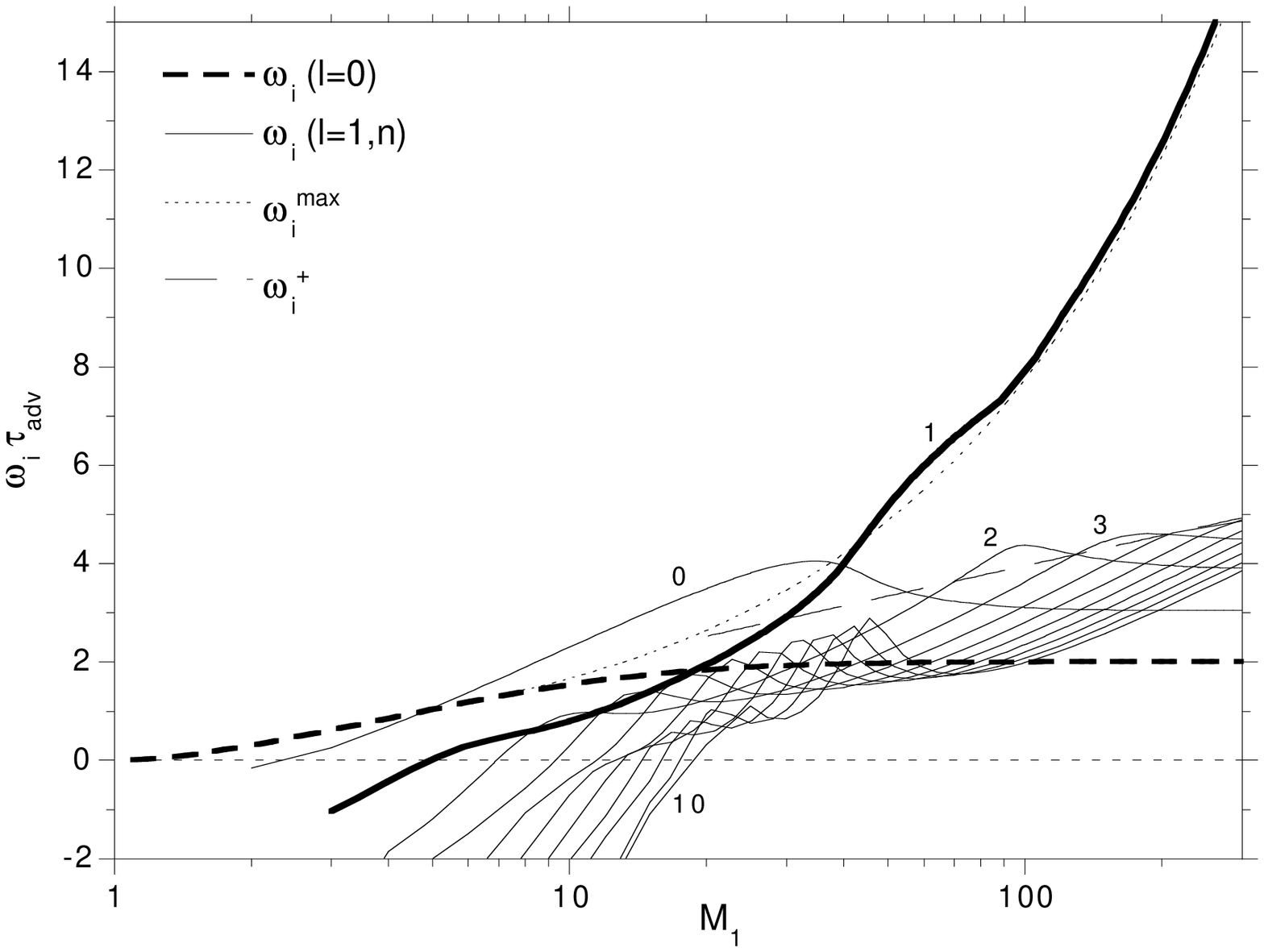,width=\columnwidth}
\psfig{file=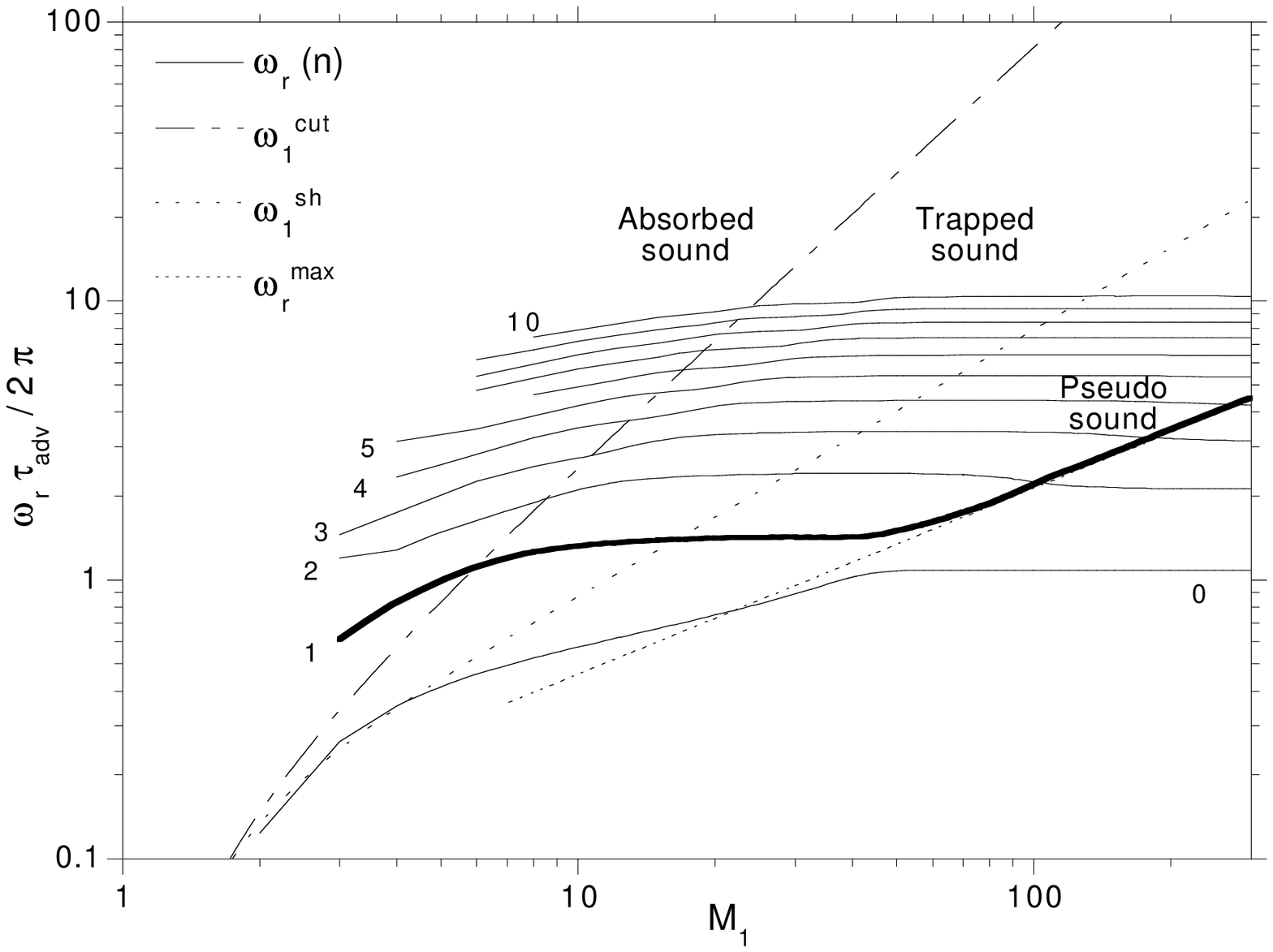,width=\columnwidth}
\caption[]{Real and imaginary parts of the eigenfrequencies of the
modes $l=1$ in the isothermal flow, for different Mach numbers.
Frequencies are multplied by the advection time. The strong $l=1$ 
instability due to postshock acceleration corresponds to the thick 
line in both plots. In the upper plot, the growth 
rate of the fastest of the mode $l=0$ is also displayed (thick dashed line). 
The analytical approximation of the fastest growth rates 
$\omega_{i}^{+},\omega_{i}^{\rm max}$ of the $l=1$
instabilities due to the entropic-acoustic cycle and to 
post-shock acceleration are shown as the thin dashed line (Eqs.~\ref{omp2})
and the thin dotted line  Eq.~(\ref{maxwi}). 
In the bottom plot, the cut-off
frequency $\omcutu$, the minimum acoustic frequency $\omega_{\rm sh}$,
and the frequency $\omega_{r}^{\rm max}$ of the most unstable mode 
(Eq.~\ref{maxwr}) are indicated as references.}
\label{figspiso}
\end{figure}
The effect of non radial perturbations ($l\ge1$) 
involves an integral on the right hand side of Eq.~(\ref{equnique}), which 
reflects the cumulative excitation of acoustic waves by the vorticity 
perturbations advected from the shock to the sonic radius.
Fig.~\ref{figM100} shows the eigenspectrum of the isothermal Bondi flow for 
$M_1=100$, $l=0,1,2,3$. This spectrum is characterized as follows:
\par (i) The most unstable mode is asymmetric ($l=1$) 
and well separated from the other eigenmodes. The growth rate of this low 
frequency mode is significantly larger than the advection time, nearly four times 
faster than the unstable radial mode. The analytical estimates of the growth 
rates at low frequency correspond to Eqs.~(\ref{wr2}-\ref{wi2}), 
(\ref{sol0r}-\ref{sol0}) and (\ref{maxwr}-\ref{maxwi})
in Sects.~\ref{Sectps} and \ref{secps}.
\par(ii) Non radial unstable modes are very numerous and cover a wide range of 
frequencies, from the advection frequency ($\sim v_{\rm sh}/\rsh$),  up to the 
cut-off frequency ($\propto c/\rso$).
\par(iii) A striking feature of the eigenspectrum in Fig.~\ref{figM100} is 
the apparent 
oscillation of the imaginary part as a function of the real part of the frequency. 
This behaviour is explained in Sect.~\ref{secvai} on the basis of the 
vortical-acoustic cycle.\\
The variation of the growth rate with the incident 
Mach number is shown in Fig.~\ref{figspiso}. The higher the Mach number, 
the more unstable the non radial modes, while the radial instability 
saturates. The growth rate of the most unstable $l=1$ mode increases 
with the incident Mach number much faster than in the other modes.
The real part of its complex frequency is intermediate 
between the advection and acoustic frequencies. It is 
computed analytically in Sect.~\ref{secps}.\\
The rest of the paper is dedicated to understanding this 
apparently complicated spectrum, and to questioning the role of post-shock 
acceleration. 

\section{The sound of vorticity\label{Svortexsound}}

\subsection{Frequency dependence of the vortical-acoustic coupling}
\begin{figure}
\psfig{file=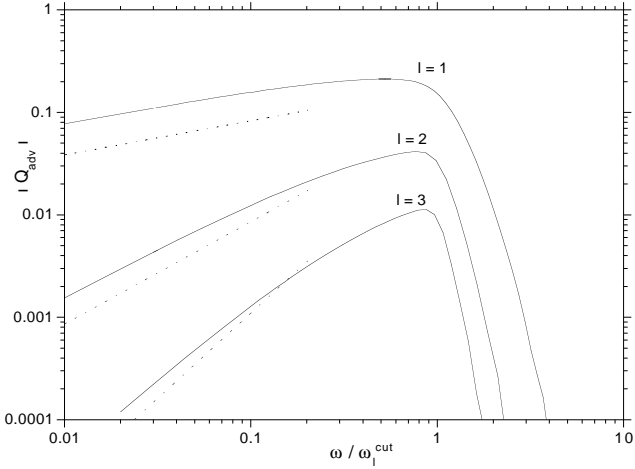,width=\columnwidth}
\caption[]{Efficiency $|{{\cal Q}}_{\rm adv}|$ of the excitation of 
acoustic waves propagating against the stream by the advection of a vortex in an
isothermal flow, for $l=1,2,3$. The frequency is in units of the
cut-off frequency $\omcutu$. The dotted lines correspond to the
asymptotic behaviour ${{\cal Q}}_{\rm adv}\propto \omega^{(2l-1)/3}$ at 
low frequency determined analytically in Eq.~(\ref{scaqk}).}
\label{figeffk}
\end{figure}
Let us assume that a vorticity wave is 
advected from infinity towards the accretor: what is the sound 
produced by the advection of this vorticity ? This question is solved 
in Appendix~D by computing the coupling coefficient 
${\cal Q}_{\rm adv}$ measuring the outgoing acoustic flux $F_{-}$ 
associated with a vorticity perturbation $\delta K$:
\begin{equation}
F_{-}={{\dot M}_{0}\over c^2}|{{\cal Q}}_{\rm adv}|^{2}
|\delta K|^{2},\label{fluxqadv}
\end{equation}
The relationship between the acoustic flux $F_{\pm}$ and the acoustic 
perturbation $f_{\pm}$ is deduced from the WKB approximation in Appendix~B:
\begin{equation}
F_{\pm}\sim 
{{\dot M}_{0}\over c^{2}}{\mu\over\M}\left|f_{\pm}\right|^{2}.\label{flux}
\end{equation}
Thus the complex efficiency ${{\cal Q}}_{\rm adv}$ is defined at a 
radius $R\gg c/\omega$, in the region of propagation of acoustic waves, 
as follows:
\begin{eqnarray}
{{\cal Q}}_{\rm adv}(R)\equiv \left({\mu\over\M}\right)^{1\over2}
{f_{-}\over\delta K },\label{defqadv}\\
=-{1\over 2i\omega c}
\e^{{i\omega\over c}\int_{\rt}^{R}{\mu-\M\over1-\M^{2}}{\d r\over \M}}
\int_{\rso}^{R} {f_{0}\over\M r^{2}}
\e^{-{i\omega\over c}\int_{R}^r{1+\M^{2}\over 1-\M^{2}}{\d r\over\M}}
\d r.\label{intqadvR}
\end{eqnarray}
${{\cal Q}}_{\rm adv}$ is approximately independent of $R$ in the WKB region 
of propagation of acoustic waves, \ie away from their turning point. 
This coefficient is approximated in Appendix~D in the low frequency limit 
$c/R\ll\omega\ll c/\rso$, \ie below the cut-off frequency, using the 
function ${\cal H}(l)$ defined from an integral of the Spherical Bessel 
function $j_{l}$ (Eq.~\ref{intcube}):
\begin{equation}
|{\cal Q}_{\rm adv}|\sim {3^{l+1\over3}\e^{2l-1\over4}
\over1\cdot3\ldots(2l+1)}
{{\cal H}(l)\over2}\left({\omega \rso\over c}\right)^{2l-1\over3},
\label{scaqk}
\end{equation}
with ${\cal H}(1)\sim 0.45$, ${\cal H}(2)\sim 0.33$ and ${\cal H}(3)\sim 
0.28$. The direct integration of Eq.~(\ref{intqadvR}) is compared to the 
approximation obtained in Eq.~(\ref{scaqk}) for $l=1,2,3$ in 
Fig.~\ref{figeffk}. The asymptotic behaviour 
$|{\cal Q}_{\rm adv}|\propto \omega^{2l-1/3}$
is correctly reproduced. The scaling factor of this power law 
obtained analytically is too low by a factor $2$ for $l=1$ perturbations, 
which can be attributed to the roughness of the approximation.
This calculation clearly indicates that the vortical-acoustic coupling is 
most efficient near the cut-off frequency. The same conclusion 
was reached in F01 concerning the efficiency of the entropic-acoustic 
coupling.

\subsection{Region of coupling\label{Sregion}}

This approximation enables us to determine the region where the coupling 
between the vorticity and the acoustic perturbations is most efficient. 
It occurs mainly in the 
subsonic region $\rso<r<r_{\rm eff}$ where the wavelength of the vorticity 
perturbation is largest (see Fig.~\ref{figpseudo}):
\begin{equation}
\omega \int_{\rso}^{r_{\rm eff}} {\d r\over v}\sim 2\pi.\label{reff}
\end{equation}
A similar result was obtained in F01 concerning entropy perturbations for 
$\gamma=5/3$. Due to the differences in velocity profiles in 
Eq.~(\ref{reff}), 
$r_{\rm eff}\propto \omega^{-2/3}$ for $\gamma=5/3$ at high frequency
(Eqs.~E.6, E.11 and E.12 of F01), whereas 
\begin{equation}
r_{\rm eff}\propto \left({\omega\rso\over 
c}\right)^{-{1\over3}}\rso\label{defreff}
\end{equation}
in the isothermal flow at low frequency. It is therefore important to note 
that the most efficient coupling at low frequency lies in the region of 
pseudosound well within the turning point ($\rt\sim c/\omega$) of 
acoustic waves:
\begin{equation} 
\rso\ll r_{\rm eff}\ll \rt.\label{inreff}
\end{equation}
According to Eq.~(\ref{reff}), the upper bound $r_{\rm eff}$ of the 
coupling region for perturbations near the advection frequency coincides with 
the shock radius.

\section{Vortical-acoustic instability at high frequency\label{secvai}}

\subsection{The vortical-acoustic cycle for $\omega_{\rm sh}\ll\omega\ll\omcut$}

The formalism developed in FT00 for the entropic-acoustic cycle within the 
WKB approximation can be transposed to the case of the 
vortical-acoustic cycle. This formalism requires the frequency to be high 
enough so that acoustic waves can be identified and their propagation time 
measured. An initial perturbation of vorticity $\delta K(t_{0},\rsh)$ at the 
shock, advected towards the accretor, triggers the excitation of an 
acoustic flux $F_{-}$ propagating outwards. As it reaches the shock 
surface, this acoustic flux induces new vorticity perturbations 
$\delta K(t_{0}+\tau_{\cal Q},\rsh)$, where $\tau_{\cal Q}$ is the duration of 
this cycle. The global efficiency ${\cal Q}$ of the cycle is naturally
\begin{equation}
{\cal Q}\equiv {\delta K(t_{0}+\tau_{\cal Q},\rsh)\over\delta K(t_{0},\rsh)}.
\end{equation}
${\cal Q}$ and  $\tau_{\cal Q}$ depend a priori on the frequency 
$\omega_{r}$ and spatial structure $l$ of the perturbation considered. 
The growth (or damping) rate of the cycle is identified with the imaginary 
part of the eigenfrequency, given by:
\begin{equation}
\omega_{i}\equiv {1\over\tau_{\cal Q}}\log |{\cal Q}|.\label{growth}
\end{equation}
The timescale $\tau_{\cal Q}$ is dominated by the 
advection timescale $\tau_{\rm adv}$ if the shock is strong 
($\M_{2}\ll1$, $\rsh\gg \rt$):
\begin{equation}
\tau_{\cal Q}\sim\int_{\rt}^{\rsh} {1\over1-\M}{\d r\over |v|} 
\sim \tau_{\rm adv}\label{tauq}.
\end{equation}
The global efficiency ${\cal Q}$ can be decomposed
in terms of the efficiencies of the coupling between 
$F_{-}$ and $\delta K$ during advection (${\cal Q}_{\rm adv}$) 
and at the shock (${\cal Q}_{\rm sh}$), defined immediately after 
the shock according to Eq.~(\ref{defqadv}) and:
\begin{eqnarray}
{{\cal Q}}_{\rm sh}&\equiv& \left({\M_{2}\over\mu}\right)^{1\over2}
{\delta K_{\rm sh} \over f_{-}},\label{defqsh}\\
{\cal Q}&= &{{\cal Q}}_{\rm adv}{{\cal Q}}_{\rm 
sh}\e^{-i\omega\tau_{\cal Q}}.\label{defQ}
\end{eqnarray}
The modulus of ${{\cal Q}}_{\rm sh}$
is directly interpreted as a coupling efficiency
between $F_{-}$ and $|\delta K|^{2}$:
\begin{equation}
|\delta K_{\rm sh}|^{2}=  {c^{2}\over{\dot M}_{0}}
|{{\cal Q}}_{\rm sh}|^{2}F_{-}.\label{fluxk}
\end{equation}
The efficiency ${{\cal Q}}_{\rm sh}$ is computed in Appendix~E, 
following a method introduced by D'Iakov (1958) and Kontorovich (1958, 1959).
${{\cal Q}}_{\rm sh}$ can be approximated 
by a WKB analysis when the wavelength of the perturbation is small 
compared to the lengthscale of the local gradients of the flow, by 
keeping only the first order terms in $\M_{2}$ and $c/\omega r$, 
at high frequency and for strong shocks:  
\begin{equation}
{{\cal Q}}_{\rm sh}\sim-{2l(l+1)\over\M_{2}^{1\over2}}
\left(1-\M_{2}- i{\eta c\over\omega r}\right).\label{qKiso2}
\end{equation}
This calculation proves that the flow reacceleration does not affect 
the vortical-acoustic coupling at the shock for
\begin{equation}
{\omega r\over c}\gg \left|\eta\right|\sim 2,
\end{equation}
and that ${{\cal Q}}_{\rm sh}$ is approximately independent of 
frequency in the range $\omega_{\rm sh}\ll\omega<\omcut$.
As a consequence of Eqs.~(\ref{defqsh}) and (\ref{qKiso2}), the 
vorticity is produced at the shock with an efficiency proportional to 
the shock strength:
\begin{equation}
\delta K_{\rm sh}\propto \M_{1} f_{-}\propto \M_{1}{\delta p_{-}\over p}.
\label{propM}
\end{equation}
The global efficiency ${\cal Q}$ should consequently peak near the
cut-off frequency, and drop abruptly above it. The frequency dependence 
of ${\cal Q}$ appears in Fig.~\ref{figQR} (top picture, full line), 
as deduced from Fig.~\ref{figeffk} and Eq.~(\ref{qKiso2}) in the WKB 
approximation. Thus the instability should be strongest near the 
cut-off frequency with the following scaling, obtained from 
Eqs.~(\ref{omcut}), (\ref{scaqk}) and (\ref{qKiso2}) for 
$\omega_{\rm sh}\ll\omega\ll\omcut$:
\begin{equation}
|{\cal Q}|\sim \left({\omega\over\omcut}\right)^{2l-1\over 3}
\left({\M_{1}\over\M_{0l}}\right)^{1\over2},\label{roughq}
\end{equation}
where the scaling factor $\M_{0l}$ depends on $l$ through both
${{\cal Q}}_{\rm adv}$ and ${{\cal Q}}_{\rm sh}$:
\begin{equation}
\M_{0l}\sim \left({4\over3\e^{3\over2}}\right)^{2l-1\over3}
{\left\lbrack 1\cdot3\ldots(2l+1)\right\rbrack^{2}\over 
\left\lbrack l(l+1)\right\rbrack^{2l+5\over3}}
{1\over3{\cal H}^{2}(l)}.
\end{equation}
$\M_{0l}\propto l^{2l/3}$ diverges for $l\gg10$, thus favouring the instability 
of low degree modes. According to Fig.~\ref{figeffk}, the 
extrapolation of Eq.~(\ref{roughq}) to $\omega\sim\omcut$ gives a very 
rough upper bound of $|{\cal Q}|(\omcut)$, particularly 
overestimlating it for high degree perturbations. Although 
$\M_{01}\sim1.96$ exceeds $\M_{02}\sim 0.95$ and $\M_{03}\sim0.69$, the actual 
efficiency $|{\cal Q}|$ is maximal for the mode $l=1$, as checked in 
Fig.~\ref{figQR} by multiplying the curve $|{\cal Q}_{\rm adv}|$ 
(Fig.~\ref{figeffk}) by $|{\cal Q}_{\rm sh}|$ (Eq.~\ref{qKiso2}).
Using Eqs.~(\ref{growth}) and (\ref{tauq}), the growth rate $\omega_{i}$ 
at high frequency is deduced from Eq.~(\ref{roughq}):
\begin{eqnarray}
&&\omega_{\rm sh}\ll\omega_{r}\ll\omcut,\\
\omega_{i}&\sim&
{1\over2\tau_{\rm adv}}\log \left({\omega\over\omcut}\right)^{2(2l-1)\over 3}
{\M_{1}\over\M_{0l}},\label{grwi}\\
&\sim&
{1\over2\tau_{\rm adv}}\log 
\left({2\omega\over\e^{3\over4}\omega_{\rm sh}}\right)^{2(2l-1)\over 3}
{\M_{1}^{2(2-l)\over3}\over\M_{0l}}.\label{grwi2}
\end{eqnarray}
Eq.~(\ref{grwi2}) indicates that the range of unstable frequencies 
below the cut-off frequency gets narrower for $l\ge2$.
This is also confirmed by the eigenspectrum obtained numerically for 
$l=1,2,3$ in Fig.~\ref{figM100}. Note that the efficiency $|{\cal Q}|$ 
was estimated using $|{{\cal Q}}_{\rm adv}|$ which was computed for
perturbations with a purely real frequency. Thus the estimate of the 
growth rate in Eq.~(\ref{grwi}) implicitly assumes that the imaginary 
part is small compared to the real part. This is true since 
$\log\M_{1}/\tau_{\rm adv}\ll \omcut$.
This calculation proves that the vortical-acoustic cycle is unstable 
for strong shocks, with a growth rate exceeding the growth rate of 
the radial mode, and a mechanism which is independent of the
sign of the local flow acceleration immediately after the shock. 
A more refined description of the entropic-acoustic cycle is 
developped in the next section 
in order to understand the apparent oscillations in the 
spectrum obtained numerically in Fig.~\ref{figM100} and 
Fig.~\ref{figspiso}.

\subsection{Additonnal contribution of the acoustic cycle}

\subsubsection{Dispersion relation for the double cycle\label{bicycle}}

The role of the purely acoustic cycle had been anticipated in FT00.
It is characterized by a time scale $\tau_{\cal R}$ and a global efficiency 
${\cal R}$. Let us show how the simultaneous existence 
of the two cycles can explain the apparent oscillation in the 
eigenspectrum of Fig.~\ref{figM100}. The following extension of the analysis 
of FT00 is valid for both entropic-acoustic and vortical-acoustic cycles, 
by replacing "vortical" by "entropic".\\ 
A perturbation $f$ is influenced by the two cycles as follows:
\begin{equation}
f(t)={\cal Q}f(t-\tau_{Q}) + {\cal R}f(t-\tau_{R}).\label{double}
\end{equation}
A solution of the form $f(t)\propto \exp(-i\omega t)$ satisfies 
Eq.~(\ref{double})  if the complex eigenfrequency ($\omega_{r},\omega_{i}$) 
is a solution of the following dispersion equation (eq.~25 of FT00):
\begin{equation}
{\cal Q}\e^{i\omega\tau_{\cal Q}}+{\cal R}\e^{i\omega\tau_{\cal R}}=1.
\label{disp}
\end{equation}
This dispersion relation is recovered in Appendix~F as a WKB 
approximation of the exact dispersion relation (\ref{equnique}).
The analysis of the dispersion relation (\ref{disp}) in Appendix~G
enables us to extract physical information from the complicated 
eigenspectrum in Fig.~\ref{figM100}, such as the ratio of timescales 
$\tau_{\cal Q}/\tau_{\cal R}$, and the dimensionless efficiencies
$|{\cal Q}|$ and $|{\cal R}|$.

\subsubsection{Ratio of the two cycles timescales $\tau_{\cal Q}/\tau_{\cal R}$}

\begin{figure}
\psfig{file=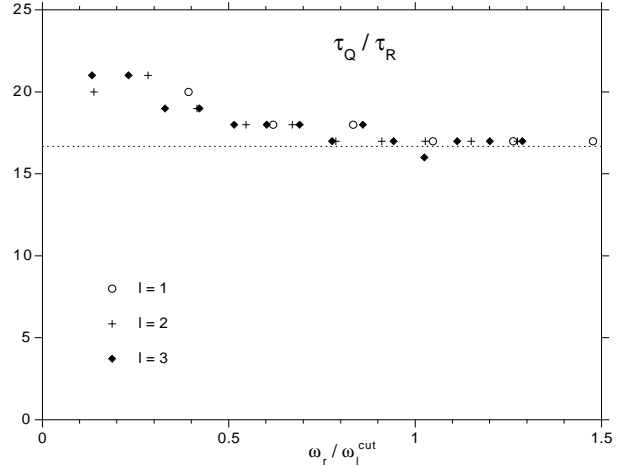,width=\columnwidth}
\caption[]{Number of eigenmodes in each oscillation of the 
eigenspectrum in Fig.~\ref{figM100}, identified as the ratio of the 
timescales of the two cycles. For strong shocks, $\tau_{\cal 
Q}/\tau_{\cal R}\sim \M_{1}/6$}
\label{fignmode}
\end{figure}
The timescale $\tau_{\cal R}$ of the acoustic cycle for strong shocks
can be written approximately:
\begin{equation}
\tau_{\cal R}\sim \int_{\rt}^{\rsh} {2\over 1-\M^{2}}{\d r\over c}\sim 
2{\rsh\over c}\propto \M_{1}^{1\over2}.\label{taur}
\end{equation}
According to Eqs.~(\ref{tauq}) and (\ref{taur}), the ratio 
$\tau_{\cal Q}/\tau_{\cal R}$ for a strong shock is simply:
\begin{equation}
{\tau_{\cal Q}\over\tau_{\cal R}}\sim{\M_{1}\over6}.
\label{scaleps}
\end{equation}
As shown in Appendix~G, this ratio is the average
number of eigenmodes in each oscillation of the eigenspectrum. 
As an illustration, this number is reported on Fig.~\ref{fignmode} for 
the modes $l=1,2,3$ in the case $\M_{1}=100$ corresponding to Fig.~\ref{figM100},
in good agreement with Eq.~(\ref{scaleps}).

\subsubsection{Efficiencies $|{\cal Q}|$ and $|{\cal R}|$}

\begin{figure}
\psfig{file=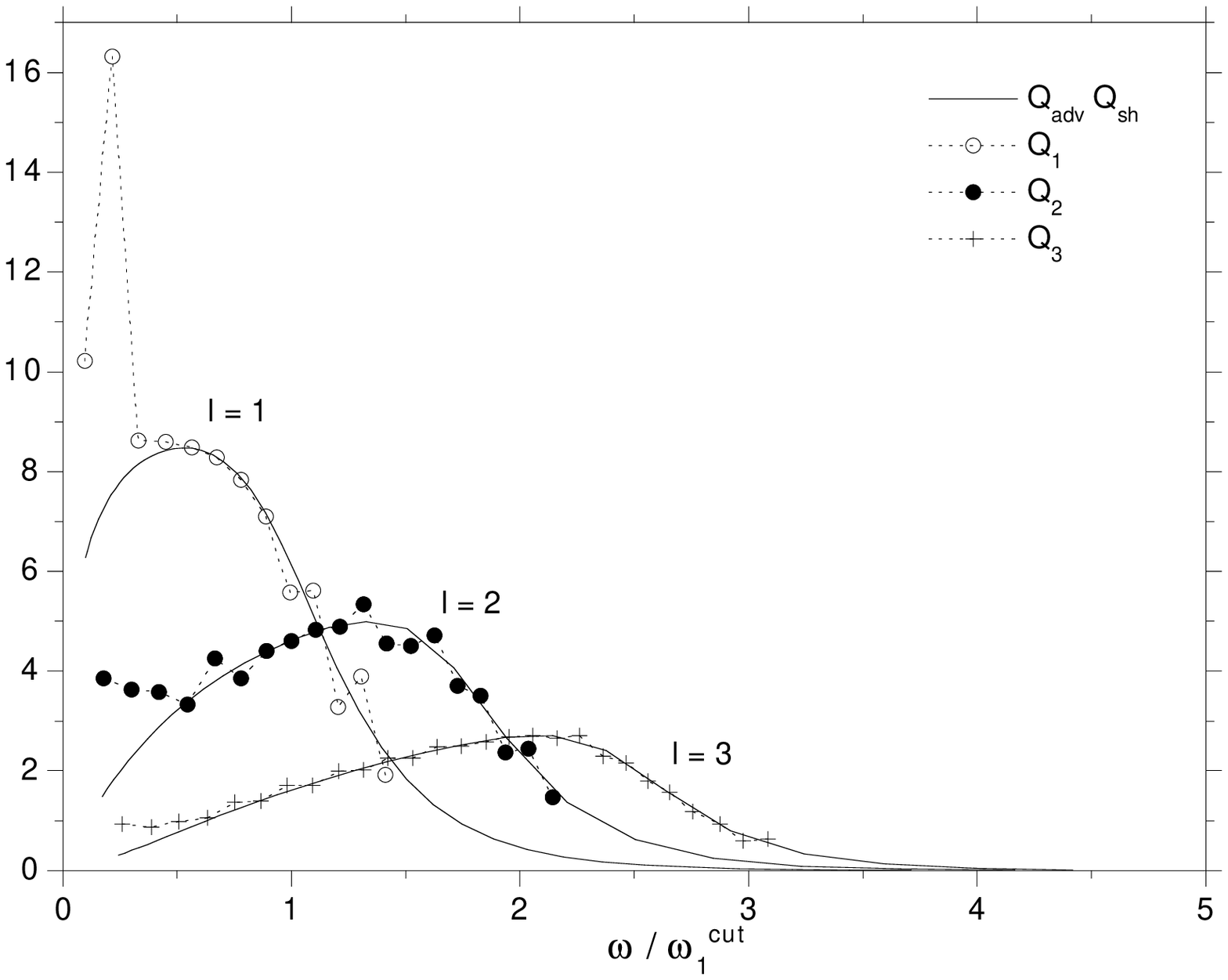,width=\columnwidth}
\psfig{file=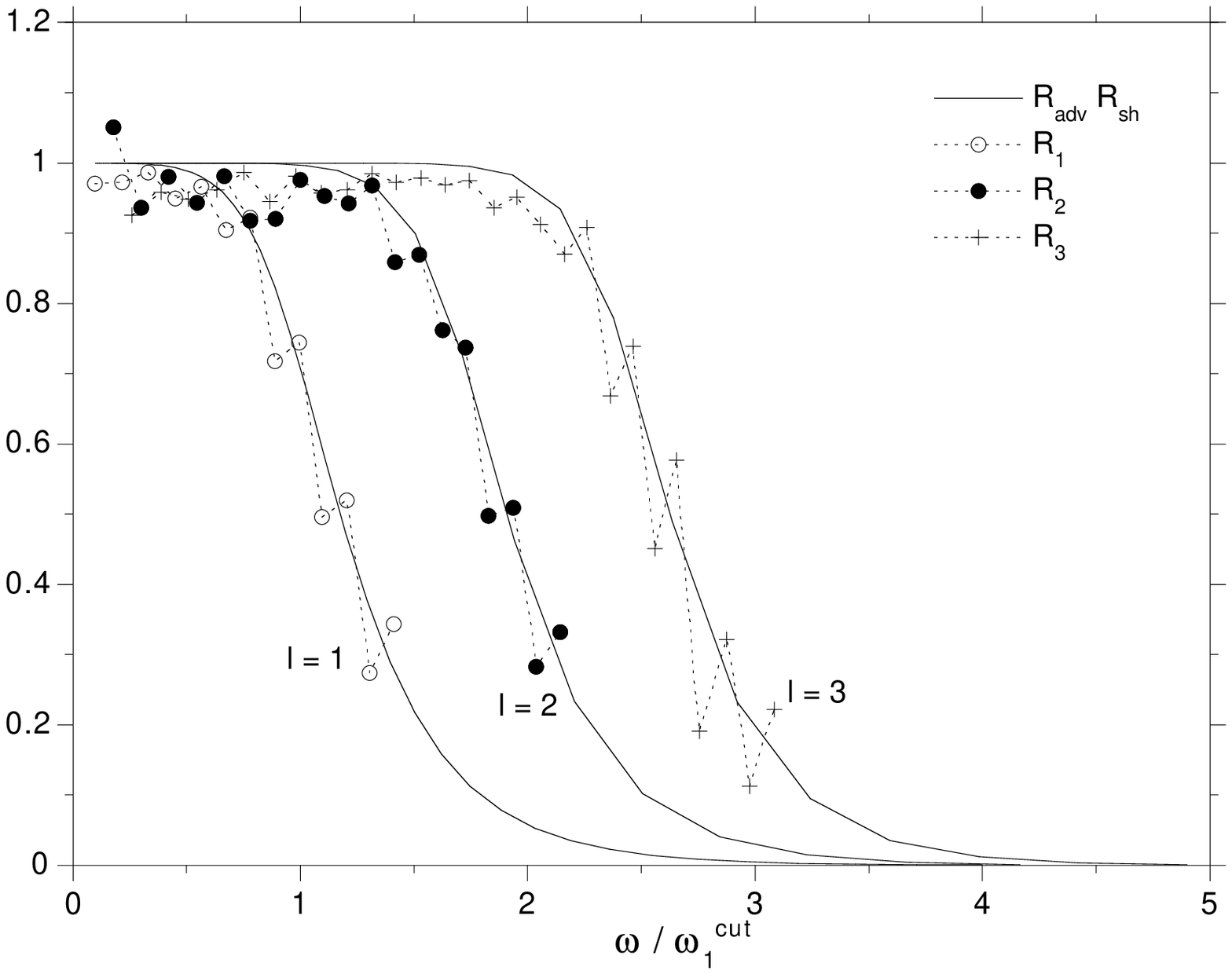,width=\columnwidth}
\caption[]{Global efficiencies $|{\cal Q}|$, $|{\cal R}|$ asociated to non radial 
perturbations $l=1,2,3$ deduced from the eigenspectrum in Fig.~\ref{figM100} 
in the framework of the double cycle model (dotted line), compared 
with the efficiencies deduced from the calculation of 
${{\cal Q}}_{\rm adv},{{\cal Q}}_{\rm sh}$ and 
${{\cal R}}_{\rm adv},{{\cal R}}_{\rm sh}$}
\label{figQR}
\end{figure}
As noticed in FT00, the acoustic cycle can be neglected ($\omega_{i}
\sim\log |{\cal Q}|/\tau_{\cal Q}$) if the efficiencies and timescales 
of the two cycles are such that $|\alpha|\ll1$, with
\begin{equation}
\alpha\equiv {{\cal R}\over{\cal Q}^{\tau_{\cal R}\over\tau_{\cal Q}}}.
\label{defalpha}
\end{equation}
If $\alpha$ is not negligible, the acoustic cycle can contribute to either 
stabilize or destabilize the vortical-acoustic cycle. The most stabilizing effect of the acoustic cycle 
is the effective reduction of ${\cal Q}$ by a factor $2$ (see Appendix~G). 
By contrast, its destabilizing contribution can be much larger 
comparatively, if the acoustic time is short compared to the advection time. 
This is the case for strong shocks. Depending on the 
relative phases of the two cycles, the growth rate can then cover the following 
range:
\begin{equation}
{1\over\tau_{\cal Q}}\log {|{\cal Q}|\over 1+|\alpha|}\le 
\omega_{i}\le {1\over\tau_{\cal Q}}\log {|{\cal Q}|\over 1-|\alpha|}.
\end{equation} 
Conversely, the values of the  dimensionless physical parameters 
$|{\cal Q}|,|{\cal R}|$ can be extracted from the eigenspectrum, 
as in Fig.~\ref{figM100}, by measuring
\par(i) the extremal values $\omega_{i}^{+},\omega_{i}^{-}$,
\par(ii) the period $\Delta\omega_{r}$ of the oscillations
\par(iii) the average number $n$ of modes per period.\\
As demonstrated in Appendix~G, the timescale of
the acoustic cycle is simply $\tau_{\cal R}=2\pi/\Delta\omega_{r}$.
Thus the oscillations are well resolved in the eigenspectrum only if the
advection time is significantly longer than the acoustic time
(\ie $\M_{1}>6$).\\
The values of $|{\cal Q}|,|{\cal R}|$ are determined using the following 
relations:
\begin{eqnarray}
|{\cal Q}|&=&{\cosh \pi{\Delta\omega_{i}\over\Delta\omega_{r}}
\over
\cosh (n-1)\pi{\Delta\omega_{i}\over\Delta\omega_{r}}}
\exp2n\pi{{\bar\omega}_{i}\over\Delta\omega_{r}},\label{calQ}\\
|{\cal R}|&=&{\sinh n\pi{\Delta\omega_{i}\over\Delta\omega_{r}}
\over
\cosh (n-1)\pi{\Delta\omega_{i}\over\Delta\omega_{r}}}
\exp2\pi{{\bar\omega}_{i}\over\Delta\omega_{r}}.\label{calR}
\end{eqnarray}
where
\begin{eqnarray}
{\bar \omega}_{i}&\equiv &{\omega_{i}^{+}+\omega_{i}^{-}\over2}\\
\Delta\omega_{i}&\equiv &\omega_{i}^{+}-\omega_{i}^{-},
\end{eqnarray}
As an illustration, Fig.~\ref{figQR} is obtained
from the eigenspectrum of Fig.~\ref{figM100} using Eqs.~(\ref{calQ}) 
and (\ref{calR}). The accuracy of the measurements of $\Delta \omega_{r},
\Delta\omega_{i}$ and ${\omega_{i}}$ is of course of the order $1/n$. 
As a check of consistency, the value of $|{\cal Q}|$ obtained  
from the product of $|{\cal Q}_{\rm adv}|$ (Fig.~\ref{figeffk}) and 
$|{\cal Q}_{\rm sh}|$ (Eq.~\ref{qKiso2}) is also displayed,
showing an excellent agreement except for the lowest frequency modes 
where WKB approximation breaks down. It is remarkable that $|{\cal Q}|$ seems 
to be maximized at low frequency for $l=1$ whereas it is maximized near the 
cut-off frequency $\omcut$ for $l\ge2$.
The efficiency $|{\cal R}|$ of the acoustic cycle is bounded by 
one and decreases to zero above the cut-off frequency, as could be expected 
from FT00 and F01.

\subsubsection{WKB approximation of ${\cal R}$ and growth rate}

The global efficiency ${\cal R}$ can be decomposed
in terms of the efficiencies ${{\cal R}}_{\rm adv}$ and 
${{\cal R}_{\rm sh}}$ of the coupling between the acoustic
fluxes $F_{\pm}$, such that the 
global efficiency ${\cal R}$ is
\begin{equation}
{\cal R}\equiv{{\cal R}_{\rm adv}}{{\cal R}_{\rm sh}}
\e^{-i\omega\tau_{\cal R}}.\label{defR}
\end{equation}
The efficiency ${{\cal R}}_{\rm adv}$ 
measures the outgoing acoustic flux $F_{-}$ produced by the 
deviation of an ingoing acoustic flux $F_{+}$:
\begin{eqnarray}
{{\cal R}_{\rm adv}}&\equiv&{f_{-}\over f_{+}},\\
&=&\rc\e^{{i\omega\over c}\int_{\rt}^{\rsh} {2\mu\over1-\M^{2}}\d r},\\
F_{-}&=&|{{\cal R}_{\rm adv}}|^{2}F_{+}.
\end{eqnarray}
${{\cal R}_{\rm adv}}$ is simply the limit when $\gamma\to1$ of the 
efficiency 
computed in F01: it is close to unity below the cut-off frequency 
$\omcut$, and decreases exponentially above this cut-off.
Conversely, an acoustic flux $F_{-}$ reaching a shock 
produces a reflected ingoing acoustic flux $F_{+}$ with the efficiency 
${{\cal R}_{\rm sh}}$:
\begin{eqnarray}
{{\cal R}_{\rm sh}}&\equiv&{f_{+}\over f_{-}}\label{defrsh},\\
F_{+}&=& |{{\cal R}_{\rm sh}}|^{2} F_{-}.\label{fluxp}
\end{eqnarray}
Keeping only the first order terms in $\M_{2}$ and $c/\omega r$, the complex 
efficiency is computed in Appendix~E:
\begin{equation}
{{\cal R}_{\rm sh}}\sim-1+2\M_{2}+ 2i{\eta c\over\omega r}.\label{riso2}
\end{equation}
Thus the global efficiency $|{\cal R}|$ should be close to unity below 
the cut-off frequency. The product $|{{\cal R}}_{\rm sh}{{\cal R}}_{\rm adv}|$ 
is displayed in Fig.~\ref{figQR} (bottom picture), in good agreement 
with the value of $|{\cal R}|$ deduced from Fig.~\ref{figM100} and 
Eq.~(\ref{calR}).\\
The extremal values $\omega_{i}^{\pm}$ of the growth rate of the 
vortical-acoustic cycle are computed 
in Appendix~G (Eq.~\ref{xpm}) for strong shocks:
\begin{equation}
\omega_{i}^{\pm}\sim{1\over\tau_{\cal Q}}\log {|{\cal Q}|\over 1\mp|\alpha|}.
\label{ompm}
\end{equation}
The asymptotic scaling of $\alpha$ is deduced from 
Eqs.~(\ref{roughq}), (\ref{scaleps}) and (\ref{riso2}) :
\begin{equation}
\alpha\sim 1-{3\over\M_{1}}\log\left\lbrack{\M_{1}\over\M_{0l}}
\left({\omega\over\omcut}\right)^{{2\over3}(2l-1)}\right\rbrack.
\end{equation}
Together with Eq.~(\ref{tauq}), and despite the roughness of the 
approximation in Eq.~(\ref{scaqk}) near the cut-off frequency, the leading 
order of the asymptotic values of $\omega_{i}^{\pm}$ is 
comparable to:
\begin{eqnarray}
\omega_{i}^{-}&\sim& {1\over2\tau_{\rm adv}}
\log {\M_{1}\over4\M_{0l}},\label{omm2}\\
\omega_{i}^{+}&\sim& {3\over2\tau_{\rm adv}}
\log {\M_{1}\over (9\M_{0l})^{1\over3}\log^{2\over3}{\M_{1}\over\M_{0l}}}
.\label{omp2}
\end{eqnarray}
The values of ${\cal Q},{\cal R}, \tau_{\cal R}/\tau_{\cal Q}$
are thus responsible for a dispersion of the growth rate by a factor $3$
in Eqs.~(\ref{omm2}-\ref{omp2}) near the cut-off frequency, depending on
the relative phases of the two cycles. This factor $3$ is consistent 
with both Figs.~\ref{figM100} and \ref{figspiso}. 

\section{Vortical-acoustic instability at low frequency\label{Sectps}}

The region of efficient coupling between the vorticity and acoustic 
perturbations lies in the region of pseudosound according to 
Eq.~(\ref{inreff}). Thus it seems natural to expect an instability at 
low frequency as long as the coupling region determined from 
Eq.~(\ref{defreff}) lies inside the shock radius, \ie down to the 
advection frequency. 
This case is similar to the hole tone instability (\eg whistling kettle) 
studied by Chanaud \& Powell (1965), or in the oscillations of impinging 
shear layers (see a review by Rockwell 1983). 
In some of these cycles, the vorticity perturbations are coupled to the 
acoustic field in the region of pseudosound. As stressed by Chanaud 
\& Powell (1965), this does not preclude the use of the term "acoustic feedback"
so that we may talk of a vortical-acoustic cycle even in this range of 
frequencies.
Although the problem cannot be treated in the WKB 
limit, the homogeneous solution can be approximated by a Spherical Bessel 
function of the first kind in the region of pseudosound far from the sonic 
point. The calculation in Appendix~H shows that the case $l=1$ and 
$l\ge2$ must be treated separately. 
In the domain of pseudosound $1\ll |\omega\tau_{\rm adv}|\ll\M_{1}$, 
Eq.~(\ref{equnique}) is reduced to:
\begin{eqnarray}
{i\omega r\over \M_{2} c}r^{2}{\p^{2} f_{0}\over\p r^{2}}\sim
&-&l(l+1)(l+\eta-2)f_{0}\nonumber\\
&+&9\;l\;\e^{i\omega \tau_{\rm adv}}
\Gamma\left({l+4\over3}\right)(i\omega\tau_{\rm adv})^{5-l\over3}f_{0}.
\label{equniqueps}
\end{eqnarray}

\subsection{Low frequency global cycle $l=1$}

The $l=1$ acoustic perturbation is approximated in Appendix~H 
as $f_{0}\propto j_{1}(\omega r/c)$, and Eq.~(\ref{equniqueps}) is 
transformed into
\begin{equation}
{9\over5}\M_{2}^{2}\left(i\omega\tau_{\rm adv}\right)^{3}={2\over3}+
\left(i\omega\tau_{\rm adv}\right)^{4\over3}\Gamma\left({5\over3}\right)
\e^{i\omega\tau_{\rm adv}}.\label{equniquel1}
\end{equation}
A branch of solutions with $\omega_{r}\gg\omega_{i}$
corresponds to a vortical-acoustic cycle between the shock and the sonic point:
\begin{eqnarray}
\omega_{i}\tau_{\rm adv}\sim 
\log\left\lbrack{3\over2}\Gamma\left({5\over3}\right)\right\rbrack
+{4\over3}\log |\omega_{r}\tau_{\rm adv}| ,\nonumber\\
{\rm if}\;\;|\omega\tau_{\rm adv}|\ll\M_{1}^{2\over3},\\
\omega_{i}\tau_{\rm adv}\sim
\log\left\lbrack{5\over9}\Gamma\left({5\over3}\right)\right\rbrack
+2\log\M_{1}-{5\over3}\log |\omega_{r}\tau_{\rm adv}| ,\nonumber\\
{\rm if}\;\;|\omega\tau_{\rm adv}|\ll\M_{1}^{2\over3}.
\end{eqnarray}
The maximum growth rate of this branch of solution is reached for a 
real frequency similar to the frequency of the local instability.
\begin{eqnarray}
\omega_{r}&\sim&{10^{1\over3}\M_{1}^{2\over3}
\over3\tau_{\rm adv}},\label{lowfwr}\\
\omega_{i}&\sim& 
{8\over 9\tau_{\rm adv}}\log{\M_{1}\over\M_{0}},\label{lowfwi}\\
\M_{0}&\equiv&{2^{5\over8}3^{3\over8}\over5^{1\over2}
\Gamma^{9\over8}\left({5\over3}\right)}\sim 1.17.
\end{eqnarray}
This growth rate is compared to the results of numerical calculations 
in Fig.~\ref{figspiso}. This growth rate exceeds the growth rate of 
the vortical-acoustic instability at high frequency (Eq.~\ref{grwi}), but 
is asymptotically smaller than the maximum growth rate $\omega_{i}^{+}$ 
reached when the purely acoustic and vortical acoustic cycles are in phase 
(Eq.~\ref{omp2}).

\subsection{Low frequency global cycle $l\ge2$}

If $l\ge2$, the acoustic feedback due to an isolated vortical perturbation near 
the shock is negligible compared to the integral effect for strong shocks. 
Eq.~(\ref{equniqueps}) is reduced to:
\begin{eqnarray}
3\Gamma\left({l+4\over3}\right)\left(i\omega\tau_{\rm adv}\right)^{2-l\over3}
\e^{i\omega\tau_{\rm adv}}=l-1
.\label{equniqueM3}
\end{eqnarray}
The most unstable modes are therefore the modes $l=2$, with
\begin{eqnarray}
\omega_{r}&\sim& {2n\pi\over\tau_{\rm adv}},
\;\;{\rm with}\;\;1\ll n \ll\M_{1},\label{wr2}\\
\omega_{i}&\sim&{\log 3\over\tau_{\rm adv}}>0.\label{wi2}
\end{eqnarray}
These approximations are in excellent agreement with the numerical 
result of Fig.~\ref{figM100}.\\
By contrast with the case of high frequency perturbations,
it seems difficult to separate completely the effect of postshock 
acceleration from the global vortical-acoustic cycle in this range of 
frequencies.

\section{Effects of post-shock acceleration: a global instability 
with a local criterion\label{secps}}

The boundary condition (\ref{boundaf}) can be rewritten 
for strong shocks as follows:
\begin{equation}
{f_{\rm sh}\over c^{2}}\sim \eta{\Delta\zeta\over \rsh}\;\;
{\rm for}\;\;|\omega|\ll{|\eta|\over\M_{2}}{c\over \rsh}\sim 
{|\eta|\over\M_{2}^{1\over2}}{c\over \rso}.\label{fzeta}
\end{equation}
This equation is similar to the argument of Nobuta \& Hanawa (1994)
concerning the balance total pressures (thermal and dynamical) 
on both sides of the shock. Rather than treating the shock surface 
as a material surface pushed by the local pressures on both sides of it, 
Eq.~(\ref{fzeta}) is interpreted as follows: an excess of Bernoulli perturbation 
$f$ (which we may call "energy") on the subsonic side of the shock is 
associated with a displacement of the shock in the direction of increase of 
the local Mach number. If $\eta<0$ (Eq.~\ref{defzeta}), an excess of 
$f$ produces an outward displacement of the shock. This statement alone is not 
conclusive: instability occurs only if the flow is not able to evacuate this 
excess of energy. Although the displacement of the shock indeed liberates 
some potential energy locally, the instability depends 
on the leakage of energy through the sonic radius (Nakayama 1993).

\subsection{Asymptotic growth rate of the radial instability}

For radial perturbations ($l=0$), the only way of evacuating the 
excess of energy is through acoustic perturbations.
Acoustic energy, however, is trapped in the subsonic region if the real 
frequency of the mode is low enough. Thus a low frequency instability 
is expected. Although the criterion of the instability is indeed local,
its growth rate depends on the spatial structure of 
the acoustic perturbation from the sonic radius to the shock, as illustrated 
by Eq.~(\ref{equnique}) for strong shocks:
\begin{equation}
{\p\log f_{0}\over\p\log r}(\rsh)\sim{1\over\eta}
\left({\omega\rsh\over c}\right)^{2}
\left(1+{i\eta c\M_{2}\over\omega\rsh}\right).\label{dlogf}
\end{equation}
$f_{0}$ is approximated as the acoustic perturbation of a uniform 
medium in spherical coordinates (see Appendix~B), using a Spherical Bessel 
function $f_{0}\propto j_{0}(\omega r/ c)$ 
for $\rsh\gg\rso$. The solution of the dispersion relation (\ref{dlogf}) 
at low frequency is a purely imaginary eigenfrequency:
\begin{eqnarray}
\omega_{r}&\sim&0,\label{sol0r}\\
\omega_{i} &\sim& {-\eta\over 3+\eta}{3\M_{2}c\over\rsh},\label{sol1}\\
&\sim & -6{\p | v|\over\p r}\sim {2\over\tau_{\rm adv}}.\label{sol0}
\end{eqnarray}
The fact that the growth time scales like the advection time is not 
obvious a priori.

\subsection{Asymptotic growth rate of the $l=1$ instability}

\begin{figure}
\psfig{file=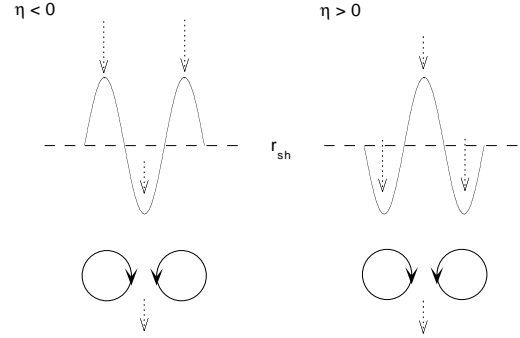,width=\columnwidth}
\caption[]{Interaction of the vortex with the shock. The contribution of a 
vorticity perturbation $\delta K_{\rm sh}>0$ to the Bernoulli perturbation is 
positive ($v\delta v_{r}>0$), and induces an increase of the local accretion 
rate ($g>0$). depending on the sign of the postshock acceleration, a 
vorticity perturbation contributes to a local expansion ($\eta>0$) or 
collapse ($\eta<0$) of the shock. }
\label{figpostshock}
\end{figure}
Non radial perturbations generate vorticity, which contribute to the 
energy balance in the subsonic part of the flow. Fig.~\ref{figpostshock} 
illustrates the contribution of a vorticity perturbation to the 
Bernoulli perturbation through the radial component of velocity 
$v\delta v_{r}$. Thus the vorticity perturbation participates to 
increase the local accretion rate (\ie $g>0$) in the regions where the shock 
moves inward ($\Delta\zeta<0$). Comparing this statement with 
$g\sim i\omega \Delta \zeta/\M_{2}c$ deduced from Eq.~(\ref{boundag}), we 
conclude that the vorticity contributes to the instability if $\nu<0$. 
A quantitative calculation of the growth rate requires taking into 
account the acoustic perturbation in the region from the shock to the sonic 
point. The fastest instability of the flow, well isolated from all the other 
modes of instability, is obtained by neglecting the last term on the right 
hand side of Eq.~(\ref{equniquel1}):
\begin{eqnarray}
\omega_{r}^{\rm max}&=&{5^{1\over3}3^{1\over2}\over2^{2\over3}}
{c\over\rsh}{1\over\M_{1}^{1\over3}},\label{maxwr}\\
\omega_{i}^{\rm max}&=&{5^{1\over3}\over2^{2\over3}}
{c\over\rsh}{1\over\M_{1}^{1\over3}}.\label{maxwi}
\end{eqnarray}
It is in excellent agreement with the numerical calculations in 
Figs.~\ref{figM100} and \ref{figspiso}.
This growth rate is intermediate between the advection and the 
fundamental acoustic frequencies. The real and imaginary parts being 
comparable, the growth rate is the time taken by the vorticity 
perturbation to travel by one wavelength $2\pi 
v/\omega_{r}^{\rm max}\propto \M_{2}^{2\over 3}\rsh\ll\rsh$. 

\section{Discussion\label{Sdiscuss}}

\subsection{Relationship with other shock instabilities}

Most shock instabilities identified and studied in astrophysics are 
related to the acceleration or deceleration of the shock itself. The 
most famous is of course the Rayleigh-Taylor instability of accelerated shocks,
analysed by  Bernstein \& Book (1978). Decelerated shocks are also 
unstable to a rippling instability, studied by Vishniac (1983), Bertshinger 
(1986) and Vishniac \& Ryu (1989). By definition, stationary shocks are 
stable with respect to these mechanisms, but can still be unstable.
Nakayama (1992, 1993) pointed out the radial instability of the shock if the 
flow is immediately accelerated after the shock, in isothermal flows. 
The validity of the postshock acceleration criterion for adiabatic flows 
is still uncertain, even for radial perturbations (Nakayama 1994).
The vortical-acoustic cycle studied in the present paper resembles 
by many aspects to the entropic-acoustic cycle studied by FT00 and F01 
in adiabatic flows. Two distinct mechanisms, however, are at work: 
\par (i) the vortical-acoustic cycle is fed by the vorticity production 
by a perturbed shock, which is highest for isothermal flows with a 
strong shock,
\par(ii) the entropic-acoustic cycle is fed by the temperature 
increase from the shock to the sonic radius, which is highest if $\gamma=5/3$
(FT00, F01).\\
A calculation similar to Appendix~E for adiabatic flows would show 
that the vorticity production by a perturbed shock is large only in the 
isothermal limit: 
\begin{equation}
\left|{\delta K_{\rm sh}\over f_{-}}\right|\propto {1\over\M_{2}}< 
\left({2\gamma\over\gamma-1}\right)^{1\over2}.
\end{equation}
Although shocked adiabatic flows with $1<\gamma<5/3$ are subject to both 
entropic-acoustic and vortical-acoustic cycles in principle, their stability 
cannot be determined without a specific calculation.

\subsection{Constraints on the instability mechanism of BHL accretion}

The stability of BHL accretion can be addressed with new tools, using
the present results and the entropic-acoustic cycle of FT00 and F01. 
Although a detailed analysis of the existing numerical simulations of 
BHL accretion is postponed to a forthcoming paper, let us outline the 
possible consequences of the entropic-acoustic and vortical-acoustic 
mechanisms on this specific flow.
Numerical simulations of BHL accretion in 3-D show a strong instability 
for adiabatic flows with $\gamma=5/3$ and small accretors 
(Ruffert \& Arnett 1994, Ruffert 1994): this coincides nicely with the 
properties of the entropic-acoustic cycle, 
which is most unstable if the temperature gradient in the subsonic part of 
the flow is strongest (FT00, F01). By contrast, nearly isothermal flows are rather 
stable in 3-D numerical simulations (Ruffert 1996). This could seem puzzling 
given the strong vortical-acoustic instability described in the present 
paper: why would the entropic-acoustic cycle be relevant for 
the BHL instability with $\gamma=5/3$, and the 
vortical-acoustic cycle be irrelevant to the BHL stability for $\gamma=1$ ?
This apparent contradiction can be partly solved by remembering the 
difference of topology between the spherical Bondi flow and the BHL 
flow, following the remark of Sect.~\ref{sectprop}. 
All the numerical simulations of isothermal BHL 
accretion seem to agree with the fact that the shock is attached to the accretor, 
whereas it is detached in high resolution simulations of adiabatic flows 
with $\gamma\ge4/3$. This can be understood qualitatively in terms of 
the weakness of pressure forces in isothermal flows compared to 
adiabatic flows ($P\propto\rho^\gamma$). More quantitatively, 
Foglizzo \& Ruffert (1997) proved that the shock in the isothermal BHL 
flow cannot be detached unless the sonic surface extends up to distances 
comparable to the Bondi radius $GM/c^{2}$, \ie much larger than the 
accretion radius. In view of this topological difference between 
isothermal and adiabatic BHL flows, it seems easier to extend the 
results obtained for shocked Bondi accretion to the subsonic region 
ahead of the accretor, for detached shocks, than in the very 
non-radial region of accretion behind the accretor for attached 
shocks. Although true, however, this argument is not conclusive. Even in a non 
radial isothermal flow, a small pressure perturbation of the shock is 
able to generate vorticity perturbations very efficiently. The lack of 
instability of isothermal BHL accretion in 3-D must be sought for in the 
lack of acoustic feedback from the advected vorticity perturbation.
Unfortunately we do not have quantitative arguments to explain why 
the acoustic feedback is so weak, apart from noticing that geometric 
compression of the vorticity perturbation from the shock to the sonic 
radius might be insufficient in the BHL flow. Any coherent explanation 
should of course also account for the strong instability observed in 2-D 
simulations of BHL accretion (Shima \etal 1998).

\subsection{Relationship with the numerical simulations of shocked inviscid 
flows with low angular momentum}

Inviscid accretion flows with low angular momentum are much more 
complex than the Bondi flow in the sense that the shock position is 
not unique (Fukue 1987, Chakrabarti 1989).
The numerical simulations of Nobuta \& Hanawa (1994) did not detect 
the effects of vorticity perturbations since they were restricted 
to axisymmetric motion. These simulations showed 
the evolution of the radial instability due to postshock 
acceleration: the unstable shock is either absorbed by the accretor 
or inflates to reach the stable outer position.\\
Molteni, T\'oth \& Kuznetsov (1999) performed non axisymmetric numerical 
simulations of shocked inviscid accretion flows with low angular momentum, 
for $\gamma=4/3$. Unstable $m=1$ oscillations were observed, with no 
physical explanation. The entropic-acoustic and vortical-acoustic 
cycles might provide a physical basis to understand this instability.

\subsection{Non linear evolution of the instability}

If the existence of the shock is postulated a priori, as in the 
present study, the flow might simply converge towards another 
solution, as illustrated by the dashed lines in Fig.~\ref{figsuper}. The shock 
can either be absorbed by the accretor, leading to the fully supersonic 
solution, or could expand towards infinity to establish the Bondi 
solution if the outer boundary condition allows it. 

In accretion flows stable with respect to the postshock acceleration 
criterion, but unstable through the vortical-acoustic (or 
entropic-acoustic) cycle, the instability might be saturated by the effect of 
the geometric dilution of the acoustic energy in the 
subsonic cavity. Let us assume that advected vorticity perturbations 
generate an acoustic flux propagating outward 
$F_{-}\propto |{\cal Q}_{\rm adv}|^{2}|\delta K_{\rm sh}|^{2}$ 
(Eq.~\ref{fluxqadv}), and that the acoustic perturbation produces in turn a 
vorticity perturbation $|\delta K_{\rm sh}|\propto \M_{1}|\delta p_{-}/p|$ 
(Eq.~\ref{propM}). The amplitude of the pressure perturbation 
$|\delta p_{-}/p|$ depends on the volume in which the acoustic flux $F_{-}$ 
is diluted through Eq.~(\ref{flux}): $|\delta p_{-}/p|\propto 
(\rsh/r) F_{-}^{1/2}$. Thus the vortical-acoustic cycle is naturally 
stabilized when the shock reaches a distance $r_{\rm max}$ defined by
\begin{equation}
r_{\rm max}\sim |{\cal Q}|\rsh,
\end{equation}
where $|{\cal Q}|$ is the global efficiency of the vortical-acoustic 
cycle in the linear regime. If the shock is not simply absorbed by 
the accretor, the non linear evolution of the instability could 
lead to quasi periodic oscillations of amplitude comparable to $r_{\rm 
max}$. However, given the number of unstable modes in a spectrum like 
Fig.~\ref{figM100}, the vortical-acoustic instability might as well 
saturate into turbulence rather than be dominated by a single QPO.
This issue can only be solved with numerical simulations.

\section{Conclusions \label{Sconclusion}}

The linear stability of shocked isothermal Bondi accretion has been 
studied by comparing the complex eigenfrequencies obtained through
a direct numerical integration (Sect.~\ref{sect3}) to the analytical 
results obtained for strong shocks by two different methods:
\par (i) an analytical estimate of the growth rate corresponding to 
a cycle of perturbations with a purely real frequency, obtained
by separating the effects of advection from the boundary effects of 
the shock. This WKB approximation, valid in the range 
of acoustic waves ($c/\rsh\ll\omega\ll c/\rso$), was used in 
Sects.~\ref{Svortexsound} and \ref{secvai},
\par (ii) an analytical estimate of the complex eigenfrequencies in the range 
of pseudosound ($v_{\rm sh}/\rsh\ll\omega\ll c/\rsh$) using Spherical Bessel 
functions (Sects.~\ref{Sectps} and \ref{secps}).\\
The results obtained by these methods are summarized as follows:

- As expected by the postshock acceleration of Nakayama (1993), the 
isothermal Bondi accretion with a shock is unstable with respect to 
radial perturbations. Its growth rate is comparable to the advection time 
from the shock to the sonic point.

- The present analysis has revealed the existence of a new instability, 
based on the cycle of vortical and acoustic perturbations in the subsonic 
part of the flow.
The analytical study of this instability at high frequency 
($c/\rsh\ll\omega\ll c/\rso$) proves 
that it is independent of the postshock acceleration criterion 
established by Nakayama (1992, 1993). It is fed by the efficient 
production of vorticity perturbations when the shock is perturbed non 
radially (Eq.~\ref{propM}), and by the 
vortical-acoustic coupling in the region of the sonic radius which 
enables the acoustic feedback. In this sense this non radial instability is 
generic and can be expected in more complex situations such as
shocked flows with a weak angular momentum accreting into a black 
hole, even if the flow is decelerated immediately after the 
shock.\\
In the shocked Bondi flow the vortical-acoustic instability is faster than 
the radial instability if the shock is strong, by a factor $\propto \log(\M_{1})$.\\
The vortical-acoustic instability occurs for low degree perturbations 
on a wide range of frequencies below the cut-off frequency $\sim c/\rso$.
A branch of unstable $l=1$ eigenmodes corresponds to a 
vortical-acoustic cycle in which the acoustic feedback is produced in the 
pseudosound domain ($v_{\rm sh}/\rsh<\omega<c/\rsh$). The resulting growth 
rate is comparable to the growth rate of the vortical-acoustic cycle at high 
frequency.\\
The role of the purely acoustic cycle was pointed out at high 
frequency in order to explain the large variations of the growth rate from 
one mode to another. More generally, the formalism developped in Appendix~G 
concerning the simultaneous acoustic and vortical-acoustic cycles applies to 
any context where the efficiencies ${\cal Q},{\cal R}$ and timescales 
$\tau_{\cal Q},\tau_{\cal R}$ can be defined.

- A strong $l=1$ oscillatory instability was found in the pseudosound domain, 
at a frequency which is intermediate between the acoustic and advection 
frequencies ($\omega_{i}^{\rm max}\propto \M_{1}^{1/3}c/\rsh$).
With comparable real and imaginary parts of the eigenfrequency, vorticity 
perturbations are advected over a very short distance during one growth time. 
On the basis of the contribution of a vortex to the Bernoulli perturbation 
sketched in Fig.~\ref{figpostshock}, this strong instability is a non radial 
consequence of post-shock acceleration.

- As outlined in Sect.~\ref{Sdiscuss}, the vortical-acoustic 
mechanism can be used as a tool in order to analyse the 
instability observed in numerical simulations of more complicated accretion 
flows. The specific application to BHL accretion or shocked flows with low 
angular momentum will be developped in a forthcoming paper.

\appendix

\section{Vorticity perturbations produced by the perturbed shock}

The non radial perturbation of velocity is deduced from the continuity 
of the velocity parallel to the shock, as in Landau \& Lifshitz (1987, 
Chapter 90, p. 336):
\begin{eqnarray}
\delta v_{\theta}&=&{v_{1}-v_{2}\over r}{\p\Delta\zeta\over\p\theta},
\label{vthet}\\
\delta v_{\varphi}&=&{v_{1}-v_{2}\over r\sin\theta}
{\p\Delta\zeta\over\p\varphi}.\label{vphi}
\end{eqnarray}
The perturbed vorticity in the flow is deduced from the 
Eqs.~(\ref{vthet}-\ref{vphi}) and the Euler equation:
\begin{eqnarray}
w_{r}&=&0,\\
w_{\theta}&=&(1-\M_{2}^{2})^{2}
\left(1-{i\eta c\M_{2}\over\omega\rsh}\right){\M_{1}^{2}\over 
r\sin\theta}{\p\Delta v\over\p\varphi}\label{wti},\\
w_{\varphi}&=&-(1-\M_{2}^{2})^{2}
\left(1-{i\eta c\M_{2}\over\omega\rsh}\right){\M_{1}^{2}\over r}
{\p\Delta v\over\p\theta}\label{wpi}.
\end{eqnarray}
The conserved quantity $\delta K_{\rm sh}$ defined by Eq.~(\ref{defKi}) 
is deduced from Eqs.~(\ref{wti}-\ref{wpi}), resulting in Eq.~(\ref{eqKsh}).

\section{Approximations of the acoustic perturbation}

\subsubsection{High frequency limit: WKB approximation}

If the wavelength of the perturbation is shorter than the lengthscale 
of the flow inhomogeneity ($\omega r/c\gg1$), a WKB approximation enables 
us to describe the propagation of acoustic waves in the direction of the 
flow ($f_{0}^{+}$) or against it ($f_{0}^{+}$):
\begin{equation}
f_{0}^{\pm}\sim {\M^{1\over2}c^{2}\over \mu^{1\over2}}\exp
i\omega\int_{\rt}^r {\M\mp\mu\over1-\M^{2}}{\d r\over c},\label{normwkb}
\end{equation}
where we have chosen the normalization of $f_{0}^{\pm}$ such that the 
lower bound of the integral is the turning point $\rt$.
The condition that $\rt<\rsh$ defines a minimum frequency 
$\omega_{l\ge1}^{\rm sh}$ using Eq.~(\ref{defomegal}) and 
$L^{2}\equiv l(l+1)$:
\begin{equation}
\omega_{l\ge1}^{\rm sh}\equiv 
\omega_{l}(\rsh)=L{c_{\rm sh}\over \rsh}
(1-\M_{\rm sh}^{2})^{1\over2}.\label{wsh}
\end{equation}

\subsubsection{Low frequency limit: uniform steady medium approximation}

Far from the accretor, the flow velocity decreases like $\propto 1/r^{2}$ 
and the density of the gas is uniform. If the shock is strong 
($\rsh\gg\rso$), the homogeneous equation associated with 
Eq.~(\ref{secordre}) is approximated for $r\gg\rso$ and 
$\omega\ll c/(r\M)$ as follows:
\begin{equation}
{\p^{2} f\over\p r^{2}}+{2\over r}{\p f\over\p r}
+\left({\omega^{2}\over c^{2}}-{L^{2}\over r^{2}}\right) f=0,
\end{equation}
which is nothing more than the equation of acoustic waves in a uniform 
steady medium, in spherical coordinates.
The solution $f_{0}$ can therefore be approximated with a 
spherical Bessel function of the first kind $j_{l}$, which is normalized 
here as in Eq.~(\ref{normwkb}):
\begin{eqnarray}
 f_{0}&\sim& e^{3\over 4} c^{2}{\omega\rso\over c}
 \left({\pi c\over 2\omega r}\right)^{1\over 2}
J_{l+{1\over2}}\left({\omega r\over c}\right),\\
&\equiv &
e^{3\over 4}c^{2}{\omega\rso\over c} j_{l}\left({\omega r\over c}\right).
\label{fbessel}
\end{eqnarray}

\section{Formulation of the boundary value problem}

The differential system (\ref{dfdri}-\ref{dgdri}) is written as a single 
differential equation of second order:
\begin{eqnarray}
{\p^{2} f\over\p r^{2}}+\left\lbrack
{\p\log\over\p r}\left({1-\M^{2}\over\M}\right)-{2i\omega \M\over 
(1-\M^{2})c}\right\rbrack{\p f\over\p r}\nonumber\\
+{{\omega^{2}\over c^{2}}-{L^{2}\over r^{2}}\over 1-\M^{2}}f=
-{\delta K_{\rm sh}\over (1-\M^{2})r^{2}}
\e^{i\omega\int_{\rsh}^r{\d r\over v}}.\label{secordre}
\end{eqnarray}
Following the method used in F01, the general solution of Eq.~(\ref{secordre}) 
can be written using the solutions $f_{0},f_{1}$ of the homogeneous equation, 
where $f_{0}$ is the unique solution which is regular at the sonic radius. 
Let us normalize $f_{0}$ using the following definition of
the WKB solutions $f_{0}^{\pm}$ (Eq.~\ref{normwkb}). 
The complex coefficient $\rc$ is defined by:
\begin{equation}
f_{0}\equiv f_{0}^{+}+\rc f_{0}^{-}.
\end{equation}
The wronskien of the couple of solutions $f_{0}^{+},f_{0}^{-}$ is deduced 
from Eqs.~(\ref{normwkb}) and (\ref{secordre}):
\begin{equation}
{\p f_{0}^{+}\over\p r}f_{0}^{-}-{\p f_{0}^{-}\over\p r}f_{0}^{+}=
-{2i\omega\M c^{3}\over1-\M^{2}}
\e^{{2i\omega\over c}\int_{\rt}^r{\M\over 1-\M^{2}}\d 
r}.\label{wronskien}
\end{equation}
Let us normalize $f_{1}$ such that the wronskien of the couple of solutions 
$f_{0},f_{1}$ is the same as in Eq.~(\ref{wronskien}). The general 
solution is then:
\begin{eqnarray}
f&=&{\delta K_{\rm sh}\over 2i\omega c^{3}}
\e^{-i\omega\int_{\rt}^{\rsh}{\d r\over v}}\nonumber\\
&&\left\lbrace
f_{0}\left\lbrack A_{0}+\int_{\rsh}^r {f_{1}\over\M r^{2}}
\e^{-{i\omega\over c}\int_{\rt}^r{1+\M^{2}\over 1-\M^{2}}{\d r\over\M}
}\d r\right\rbrack\right.\nonumber\\
&-&\left.f_{1}\left\lbrack A_{1}+\int_{\rso}^r {f_{0}\over\M r^{2}}
\e^{-{i\omega\over c}\int_{\rt}^r{1+\M^{2}\over 1-\M^{2}}{\d r\over\M}
}\d r\right\rbrack\right\rbrace.\label{solglob1}
\end{eqnarray}
A Frobenius analysis of $f_{0},f_{1}$ near the sonic points leads to:
\begin{eqnarray}
f_{0}&\sim& f_{0}(\rsh) + {\cal O}(r-\rso),\\
f_{1}&\propto& (r-\rso)^{-{i\omega\over {\dot\M}c}}.
\end{eqnarray}
thus the integrals in Eq.~(\ref{solglob1}) are converging when $r\to\rso$
if $\omega_{i}>c(\p\M/\p r)(\rso)$. The regularity at the sonic radius
therefore requires $A_{1}=0$. 
A combination of Eq.~(\ref{solglob1}) and its derivative at the shock 
radius leads to eliminate $A_{0}$ as follows:
\begin{eqnarray}
\left({\p f_{0}\over\p r}f-{\p f\over\p r}f_{0}\right)_{\rsh}=\nonumber\\
{\M_{2}\over1-\M_{2}^{2}}\delta K_{\rm sh}
\int_{\rso}^{\rsh} {f_{0}\over\M r^{2}}
\e^{-{i\omega\over c}\int_{\rsh}^r{1+\M^{2}\over 1-\M^{2}}{\d r\over\M}}
\d r.
\label{df0}
\end{eqnarray}
Eqs.~(\ref{boundaf}-\ref{boundag}) and (\ref{eqKsh}) provide the boundary 
conditions at the shock in order to replace $f(\rsh)$, $\p f/\p r (\rsh)$ and 
$\delta K_{\rm sh}$ in Eq.~(\ref{df0}), resulting in Eq.~(\ref{equnique}).

\section{Calculation of ${\cal Q}_{\rm adv}$} 

The analytical expression for ${{\cal Q}}_{\rm adv}$ can be determined 
by writting the solution corresponding to zero acoustic flux $F_{+}$ 
at an outer boundary $R$, using the couple of homogeneous solutions 
($f_{0}^{+},f_{0}^{-}$):
\begin{eqnarray}
f&=&{\delta K_{R}\over 2i\omega c^{3}}
\e^{-i\omega\int_{\rt}^{R}{\d r\over v}}\nonumber\\
&&\left\lbrace
f_{0}^{+}\left\lbrack B_{0}+\int_{\rso}^R {f_{0}^{-}\over\M r^{2}}
\e^{-{i\omega\over c}\int_{\rt}^r{1+\M^{2}\over 1-\M^{2}}{\d r\over\M}}
\d r\right\rbrack\right.\nonumber\\
&-&\left.f_{0}^{-}\left\lbrack B_{1}+\int_{\rso}^R {f_{0}^{+}\over\M r^{2}}
\e^{-{i\omega\over c}\int_{\rt}^r{1+\M^{2}\over 1-\M^{2}}{\d r\over\M}
}\d r\right\rbrack\right\rbrace.\label{solglob2}
\end{eqnarray}
The regularity at the sonic radius requires 
\begin{equation}
B_{1}=-\rc B_{0}.\label{regson}
\end{equation}
The condition of absence of an incoming acoustic flux at the outer 
boundary is obtained by canceling the coefficient of $f_{0}^{+}$ at 
$r=R$, in the WKB limit of high frequency ($\omega\gg c/R$):
\begin{equation}
B_{0}=-\int_{\rso}^{R} {f_{0}^{-}\over\M r^{2}}
\e^{-{i\omega\over c}\int_{\rt}^r{1+\M^{2}\over 1-\M^{2}}{\d r\over\M}}.
\label{noin}
\end{equation}
Together with Eqs.~(\ref{regson}) and (\ref{noin}), Eq.~(\ref{solglob2}) 
at the outer boundary becomes:
\begin{eqnarray}
f(R)&=&
-{\delta K_{R}\over 2i\omega c^{3}}
\e^{{i\omega\over c}\int_{\rt}^{R}{\d r\over \M}}\nonumber\\
&&f_{0}^{-}\int_{\rso}^R {f_{0}\over\M r^{2}}
\e^{-{i\omega\over c}\int_{\rt}^r{1+\M^{2}\over 1-\M^{2}}{\d r\over\M}
}\d r.\label{solglobR}
\end{eqnarray}
${{\cal Q}}_{\rm adv}(R)$, defined by Eq.~(\ref{defqadv}), is deduced 
from the asymptotic behaviour of $f_{0}^-$ in the WKB 
approximation (Eq.~\ref{normwkb}).
This calculation is formally similar to the calculation of ${\cal Q}_{K}$ in F01,
corrected for a phase shift. The efficiency ${{\cal Q}}_{\rm adv}$ involved 
in the vortical-acoustic cycle is deduced from 
Eq.~(\ref{intqadvR}), with $R=\rsh$.
In the strong shock limit, below the cut-off frequency 
($c/\rso\gg\omega\gg c/\rsh$), the acoustic efficiency ${\cal Q}_{\rm adv}$ is 
approximated using the Spherical Bessel function $j_{l}$ (\ref{fbessel}):
\begin{eqnarray}
|{\cal Q}_{\rm adv}|&\sim&{\e^{-{3\over4}}\over2}{c\over \omega \rso}
\left|\int_{0}^\infty 
\e^{-i\lambda x^{3}}j_{l}(x)\d x\right|,\label{qadv1}\\
\lambda&\equiv&{\e^{-{3\over2}}\over3}\left({c\over\omega \rso}\right)^{2}.
\end{eqnarray}
For $\lambda\gg1$, a function ${\cal H}(l)$ may be defined such that
\begin{equation}
\left|\int_{0}^\infty \e^{-i\lambda x^{3}}j_{l}(x)\d x\right|\sim 
{1\over 1\cdot3\ldots(2l+1)}
{{\cal H}(l)\over\lambda^{{l+1\over 3}}}.\label{intcube}
\end{equation}
The main contribution to the integral in 
Eq.~(\ref{intcube}) comes from the region $x\sim \lambda^{-1/3}$. 
Eq.~(\ref{scaqk}) is obtained from Eqs.~(\ref{qadv1}) and (\ref{intcube})

\section{Decomposition of the perturbation onto vorticity waves and
acoustic waves in the WKB approximation}

The perturbations $f,g$ immediately after the shock, defined by 
Eqs.~(\ref{boundaf}) and (\ref{boundag}) are decomposed as follows:
\begin{eqnarray}
f(\rsh)&=&f_{-}+f_{+}+f_{K},\label{decompf}\\
g(\rsh)&=&g_{-}+g_{+}+g_{K},\label{decompg}
\end{eqnarray}
where $f_{K},g_{K}$ correspond to the vorticity wave 
associated with the vorticity perturbation $\delta K$, and 
$f_{\pm},g_{\pm}$ correspond to the purely acoustic waves propagating 
in the direction of the flow (index $+$) or against the flow 
(index $-$).
An exact calculation can be made in the case of the reflexion 
of an acoustic wave $\delta p_{-}$ with wavevector 
($k_{\parallel},k_{\perp}$) on a plane shock in cartesian 
coordinates, in the absence of a gradient of $\M$.
The vorticity wave $f_K,g_K$ is advected at the velocity of the fluid:
\begin{eqnarray}
{\p f_K\over\p r}&=& {i\omega \over v}f_K,\\
{\p g_K\over\p r}&=& {i\omega \over v}g_K.
\end{eqnarray}
Replacing these derivatives in Eqs.~(\ref{dfdri}-\ref{dgdri}), we obtain:
\begin{eqnarray}
f_K&=&{\M_{2}^{2}c^{2}\delta K\over r^{2}\omega^{2}+v^{2}L^{2}},\label{fe}\\
g_K&=&{\delta K\over r^{2}\omega^{2}+v^{2}L^{2}}.\label{ge}
\end{eqnarray}
\begin{equation}
g_\pm=\pm{f_{\pm}\over\M_{2} c^{2}}.\label{gwkb}
\end{equation}
Eqs.~(\ref{fe}), (\ref{ge}) and (\ref{gwkb}) are used with 
Eqs.~(\ref{boundaf}-\ref{boundag}) in order to obtain Eq.~(\ref{eqKsh}) and
\begin{eqnarray}
f_{\pm}&=&{\M_{2} c^{2}\over2\mu}{\Delta v\over v}(1-\M_{2}^{2})
{(\mu\mp\M_{2})^{2}\over1\mp\mu\M_{2}},\label{fpm}
\end{eqnarray}
where $\mu$ in cartesian coordinates is also defined by 
Eq.~(\ref{defmu}), but replacing $l(l+1)/r^{2}$ by the wavenumber 
$k_{\perp}^{2}$.  
According to the definitions of ${{\cal Q}}_{\rm sh}$ and 
${{\cal R}_{\rm sh}}$ in Eqs.~(\ref{defqsh}) and (\ref{defrsh}), 
together with Eqs.~(\ref{eqKsh}) and (\ref{fpm}):
\begin{eqnarray}
{{\cal R}_{\rm sh}}
&=&-\left({\mu-\M_{2}\over\mu+\M_{2}}\right)^2
 \left({1+\mu\M_{2}\over1-\mu\M_{2}}\right),\label{riso}\\
{{\cal Q}}_{\rm sh}
&=&-2l(l+1)\left({\mu\over\M_{2}}\right)^{1\over2}
{(1+\mu\M_{2})(1-\M_{2}^{2})\over(\mu+\M_{2})^{2}}.\label{qkiso}
\end{eqnarray}
These equations show the decrease of the vortical-acoustic 
coupling for weak shocks ($\M_{2}\sim1$), and the existence of maximal 
efficiency at low frequency. Indeed, the maximum 
${{\cal Q}}_{\rm sh}\sim {3^{3\over2}\over8}l(l+1)\M_{1}^{2}$ 
is reached for a frequency such that $\mu\sim \M_{2}/3$, with 
${{\cal R}_{\rm sh}}\sim -1/ 4$. \\
The angle $\psi$ between 
the direction of propagation of the wave and the vector orthogonal to the 
shock surface is given by:
\begin{equation}
\tan\psi = \left\lbrack\left({\omega \over k_{\perp} c}\right)^{2}+\M^{2}-1
\right\rbrack^{-1}.
\label{defpsi}
\end{equation}
As remarked by Kontorovich (1958), the reflected sound wave propagates away 
from the shock with the same angle as the incident wave 
($\psi_+=\psi_-$), as in a classical reflexion.
The presence of a gradient of $\M_{2}$ in the Bondi flow precludes 
the use of these formulae at low frequency. In the following 
calculation we assume $\omega r/c\gg1$ and $\M_{2}\ll1$, and keep only 
the first order terms in $\M_{2}$ and $c/\omega r$. Thus $\mu\sim 1$.
Neglecting the coupling between the vorticity and acoustic waves in 
the vicinity of the shock, the vorticity wave $f_K,g_K$ is still advected 
at the velocity of the fluid. Eqs.~(\ref{fe}) and (\ref{ge}) are now 
approximated at high frequency by
\begin{eqnarray}
f_K&\sim&{\M_{2}^{2}c^{2}\delta K\over r^{2}\omega^{2}},\label{fewkb}\\
g_K&\sim&{\delta K\over r^{2}\omega^{2}}.\label{gewkb}
\end{eqnarray}
Acoustic waves are described by Eqs.~(\ref{dfdri}-\ref{dgdri}) in the 
absence of vorticity perturbations, \ie when $\delta K=0$.
Using the WKB approximation of Eq.~(\ref{normwkb}), the radial 
derivative of $f_\pm$ is approximated by:
\begin{equation}
{\p f_\pm\over\p r}\sim{i\omega\over c}
\left(\mp 1+\M_{2}-{i\eta c\over2\omega r}\right) f_\pm.
\label{dfwkb}
\end{equation}
$g_{\pm}$ is deduced from Eqs.~(\ref{dfdri}) and (\ref{dfwkb}):
\begin{equation}
g_\pm\sim\left(\pm1 +{i\eta c\over2\omega r}\right)
{f_{\pm}\over\M_{2} c^{2}}.\label{gwkb2}
\end{equation}
Eqs.~(\ref{fewkb}), (\ref{gewkb}) and (\ref{gwkb2}) are used with 
Eqs.~(\ref{boundaf}-\ref{boundag}) in order to obtain Eq.~(\ref{eqKsh}) and
\begin{eqnarray}
f_{\pm}&\sim&{\M_{2} c^{2}\over2}{\Delta v\over v}
\left(\pm1-\M_{2}- i{\eta c\over\omega r}\right),\label{fpmwkb}
\end{eqnarray}
Combining Eq.~(\ref{eqKsh}) with Eq.~(\ref{fpmwkb}), we obtain the 
expressions for ${{\cal R}}_{\rm sh},{{\cal Q}}_{\rm sh}$ in 
Eqs.~(\ref{qKiso2}) and (\ref{riso2}).
The pressure perturbations 
$\delta p_{\pm}$ associated with the acoustic waves $f_{\pm}$ are deduced 
from Eqs.~(\ref{deff}-\ref{defg}) and Eq.~(\ref{gwkb2}):
\begin{eqnarray}
{\delta p_{\pm}\over p}
&=&{f_{\pm}-v^{2}g_{\pm}\over c^{2}-v^{2}},\\
&\sim& (1\mp\M_{2}){f_{\pm}\over c^{2}}.
\label{dppmwkb}
\end{eqnarray}

\section{WKB approximation of the dispersion relation}

Using the integral expression of ${\cal Q}_{\rm adv}$ (Eq.~\ref{intqadvR}), 
Eq.~(\ref{equnique}) can be approximated as follows
for strong shocks ($\M_{2}\ll1$):
\begin{eqnarray}
\left\lbrack \eta - 
\M_{2}{i\omega \rsh\over c}\right\rbrack{c\over i\omega}{\p f_{0}\over\p r}
+\left\lbrack {i\omega \rsh\over c}-\M_{2}\eta\right\rbrack 
f_{0}\sim\nonumber\\
-{2l(l+1)\over\M_{2}^{1\over2}} 
\left\lbrack{i\omega \rsh\over c}+\M_{2}\eta\right\rbrack
{\cal Q}_{\rm adv}f_{0}^+
.\label{equniqueM2}
\end{eqnarray}
In the WKB limit $\omega \rsh/c \gg 1$, Eq.~(\ref{equniqueM2}) can be simplified 
into
\begin{eqnarray}
f_{0}^+ +\rc f_{0}^-\sim-2l(l+1){\cal Q}_{\rm adv}{f_{0}^+\over\M_{2}^{1\over2}}
.\label{equniqueWKB}
\end{eqnarray}
Using the expressions for ${\cal Q}_{\rm sh}$ and ${\cal R}_{\rm sh}$ 
(Eqs.~\ref{qKiso2} and \ref{riso2}), and the definition of ${\cal Q}$ and ${\cal 
R}$ (Eqs.~\ref{defQ} and \ref{defR}), the global dispersion relation 
(\ref{disp}) is recovered. 

\section{Analysis of the phase relation of the two cycles}

Let us introduce the new complex variable
\begin{equation}
z\equiv {\cal Q}\e^{i\omega\tau_{\cal Q}}.\label{defz}
\end{equation}
The resolution of the dispersion relation (\ref{disp}) is then equivalent to 
finding the complex number $z$ satisfying:
\begin{equation}
z+\alpha z^\epsilon=1,\label{dispcomplex}
\end{equation}
where the dimensionless parameter $\epsilon\equiv \tau_{\cal 
R}/\tau_{\cal{Q}}<1$ and $|\alpha|<1$ is defined by Eq.~(\ref{defalpha}). 
The minimum and maximum values $z_{\pm}$ of $|z|$ satisfy an equation 
similar to Eq.~(E3) of FT00:
\begin{equation}
z_{\pm}\mp |\alpha| z_{\pm}^\epsilon=1.\label{idemE3}
\end{equation}
The minimum and maximum effect of the acoustic cycle on the growth 
rate $\omega_{i}^\pm$ are directly related to $z_{\pm}$ 
through Eq.~(\ref{defz}):
\begin{equation}
\omega_{i}^{\pm}\equiv {1\over\tau_{\cal Q}}\log{|{\cal Q}|\over z_{\mp}}.
\label{defx}
\end{equation}
From Eq.~(\ref{idemE3}),
\begin{equation}
z_{+}+z_{-}=2,
\end{equation}
we deduce that the maximum stabilizing effect of the acoustic 
cycle is to divide the efficiency ${\cal Q}$ by a factor $2$:
\begin{equation}
1< z_{+}\le 2.
\end{equation}
If the acoustic time is much shorter than the advection time 
($\epsilon\ll1$), we deduce from Eq.~(\ref{idemE3}) the values of $x_{\pm}$:
\begin{eqnarray}
z_{\pm}\sim 1 \pm |\alpha|.\label{xpm}
\end{eqnarray}
Thus the acoustic cycle may participate efficiently to the instability 
if $\epsilon\ll1$ and $|\alpha|\sim 1$.
In the more general case where $0<\epsilon<1$, the following bounds on
$x$ are obtained from Eq.~(\ref{dispcomplex}):
\begin{equation}
1-|\alpha|\le |z|\le{1\over1-|\alpha|}.\label{boundz}
\end{equation}
The resolution of Eq.~(\ref{dispcomplex}) can be decomposed into a
phase condition applied to the points of a continuous curve ${\cal C}$, 
defined by:
\begin{equation}
|z-1|=|\alpha | |z|^\epsilon.\label{dispnorm}
\end{equation}
${\cal C}$ is a closed curve containing the point $(1,0)$ in its
interior, and $(0,0)$ in its exterior. It can be described in a univoque 
way by the angle $\varphi\equiv {\rm Arg}(z-1)\in[-\pi,\pi]$.
Let us define the angle $\theta\equiv {\rm Arg}(z)\in[-\pi,\pi]$.
The solutions of Eq.~(\ref{dispcomplex}) are recovered by applying to the 
solutions of Eq.~(\ref{dispnorm}) the following phase condition:
\begin{equation}
\varphi+2k\pi=\epsilon(\theta+2k'\pi),\label{disphase}
\end{equation}
where $k,k'$ are two integers. 
Since $(0,0)$ is exterior to ${\cal C}$, the range of values covered 
by $\theta$ when $\varphi$ covers $[-\pi,\pi]$ is limited to
$|\theta|\le\theta_{\rm max}<\pi$. The discrete solutions of the phase equation
(\ref{disphase}) can be seen in a graphic way as the intersection of the
periodic curve $\theta(\varphi)$ with the straight line
$\theta=\varphi/\epsilon$. The number of solutions in each phase 
$\varphi\in[0,2\pi]$ is therefore equal to $1/\epsilon$ on average.
Comparing Eq.~(\ref{defz}), Eq.~(\ref{dispnorm}) with the definitions 
of $\varphi,\theta$, one period of $\varphi$ corresponds to a
variation of $\omega_{r}$ of $2\pi/\tau_{\cal R}$, and one period of $\theta$
corresponds to $2\pi/\tau_{\cal Q}$.
The values of $|{\cal Q}|$ and $|{\cal R}|$ in Eqs.~(\ref{calQ}) and (\ref{calR})
can be determined from the
measurement of $\omega_{i}^+$, $\omega_{i}^-$ and the average number
$n=\tau_{\cal Q}/\tau_{\cal R}$ of eigenmodes per period
$\Delta\omega_{r}=2\pi/\tau_{\cal R}$, by eliminating $|\alpha|,x_{\pm}$
from the set of equations (\ref{idemE3}), (\ref{defx}), and 
(\ref{defalpha}).

\section{Pseudosound approximation of the dispersion relation}

The integral involved in Eq.~(\ref{equnique}) can be approximated in the low 
frequency limit by introducing the complex variable $z$:
\begin{equation}
{d z\over\d r}\equiv i{\omega\over \M c}{1+\M^{2}\over1-\M^{2}}.
\end{equation}
The contour of integration is deformed in the complex plane 
introducing a point on the real axis at $+\infty$, and performing two 
integrations by parts:
\begin{eqnarray}
\rsh\int_{\rso}^{\rsh}{f_{0}\over\M r^{2}}
\e^{-i{\omega\over c}\int_{\rsh}^r{1+\M^{2}\over1-\M^{2}}{\d r\over\M}}\d r=
\nonumber\\
{\rsh c\over i\omega}\e^{z_{\rm sh}}\left\lbrace
\int_{z_{\rm son}}^{+\infty}
{f_{0}\over r^{2}}{1-\M^{2}\over1+\M^{2}}\e^{-z}\d z+\right.\nonumber\\
\left. \int_{z_{\rm sh}}^{+\infty}\e^{-z}{\p^{2}\over\p z^{2}}
\left({f_{0}\over r^{2}}{1-\M^{2}\over1+\M^{2}}\right)\d z\right\rbrace
\nonumber\\
-{c\over i\omega \rsh}{1-\M^{2}\over1+\M^{2}}\left\lbrace f_{0}
+{\M c \rsh^{2}\over i\omega}{\p\over\p r}
\left({f_{0}\over r^{2}}{1-\M^{2}\over1+\M^{2}}\right)\right\rbrace_{\rsh}.
\label{approxG}
\end{eqnarray}
The turning point in the Bondi flow would be at $\rt\sim L c/\omega $.
For $r\gg\rso$, the homogeneous solution $f_{0}$ is 
approximated by a Spherical Bessel function of the first kind 
$j_{l}(\omega r/c)$. For $\omega r/c\ll 1$, it is approximated as 
follows:
\begin{equation}
f_{0}(r)\sim f_{0}(\rsh) \left({r\over \rsh}\right)^l.
\end{equation}
The first integral on the right hand side of Eq.~(\ref{approxG}) 
is approximated using a Gamma function:
\begin{eqnarray}
z&\sim& {i\omega r\over 3\M c},\\
{\rsh c\over i\omega}
\int_{z_{\rm son}}^{+\infty}
{f_{0}\over r^{2}}{1-\M^{2}\over1+\M^{2}}\e^{-z}\d z&\sim&
{f_{0}(\rsh)\over 3\M z_{\rm sh}^{l+1\over3}}
\int_{0}^{+\infty}z^{l-2\over3}\e^{-z}\d z,\\
&\sim&{f_{0}(\rsh)\over 3\M z_{\rm sh}^{l+1\over3}}
\Gamma\left({l+1\over3}\right).\label{apint}
\end{eqnarray}
The second integral on the right hand side of Eq.~(\ref{approxG}) is 
negligible if $\rt\gg\rsh\gg\rso$. 
In view of the particular case $l=1$, the differential equation 
satisfied by $f_{0}$ is used to sum up terms of same order:
\begin{eqnarray}
\eta r{\p f_{0}\over\p r} + l(l+1)f_{0} = r^{2}(1-\M^{2}){\p^{2} 
f_{0}\over\p r^{2}}+\left({\omega r\over c}\right)^{2}f_{0}\nonumber\\
-\M^{2}\left(\eta+{2i\omega r\over c\M}\right) r{\p f_{0}\over\p r}.
\label{dif0}
\end{eqnarray}
Using Eqs.~(\ref{approxG}), (\ref{apint}) and (\ref{dif0}), the 
dispersion relation (\ref{equnique}) is approximated in the 
pseudosound domain as follows:
\begin{eqnarray}
r^{2}{\p^{2}f_{0}\over\p r^{2}}
+ {l(l+1)\over 3z_{\rm sh}} \left\lbrack r{\p f_{0}\over\p r} +
f_{0}{\p\log\over\p\log r}\left({\M\over r^{2}}\right)\right\rbrack
\nonumber\\
\sim 3l {f_{0}\e^{z_{\rm sh}}\over z_{\rm sh}^{l-2\over3}}
\Gamma\left({l+4\over3}\right).\label{eqpseudo}
\end{eqnarray}
The left hand side of Eq.~(\ref{eqpseudo}) is dominated 
by the second derivative of $f_{0}$ if $l\ge2$:
\begin{eqnarray}
{l-1\over3}z_{\rm sh}^{l-2\over3}\e^{-z_{\rm sh}} \sim 
\Gamma\left({l+4\over3}\right).\label{l2}
\end{eqnarray}
If $l=1$, the second derivative of $f_{0}$ is approximated as follows:
\begin{eqnarray}
j_{1}(x)&\sim&{x\over3},\\
x^{2}{\p^{2}j_{1}\over\p x^{2}}&\sim& -{x^{3}\over 5} .
\end{eqnarray}
Thus Eq.~(\ref{eqpseudo}) becomes:
\begin{eqnarray}
{9\over 5} \M^{2}z_{\rm sh}^{3}
+ {2\over 9} \left\lbrack 1 +
{\p\log\over\p\log r}\left({\M\over r^{2}}\right)\right\rbrack
\sim z_{\rm sh}^{4\over3} \e^{z_{\rm sh}}
\Gamma\left({5\over3}\right).\label{l1}
\end{eqnarray}


\begin{thebibliography}{}
\bibitem[]{b78}
Bernstein I.B., Book D.L., 1978, ApJ 225, 633

\bibitem[]{b86}
Bertschinger E., 1986, ApJ 304, 154

\bibitem[]{c89}
Chakrabarti S.K., 1989, MNRAS 240, 7

\bibitem[]{cp65}
Chanaud R.C., Powell A., 1965, J. Acoust. Soc. Am. 37, 902

\bibitem[]{d58}
D'Iakov S.P., 1958, Soviet Physics JETP 6, 729, 739

\bibitem[]{fw69}
Ffowcs Williams J.E., 1969, Ann. Rev. Fluid. Mech. 1, 197

\bibitem[1997]{fr97}
Foglizzo T., Ruffert M., 1997, A\&A 320, 342

\bibitem[]{ft00}
Foglizzo T., Tagger M., 2000, A\&A 363, 174 (FT00)

\bibitem[]{f01}
Foglizzo T., 2001, A\&A 368, 311 (F01)

\bibitem[]{f87}
Fukue J., 1987, PASJ 39, 309

\bibitem[]{h75}
Howe M.S., 1975, J. Fluid Mech. 71, 625

\bibitem[]{hoy39}
Hoyle F., Lyttleton R.A., 1939, Proc. Cam. Phil. Soc.~35, 405

\bibitem[1993]{ishii:etal993}
Ishii T., Matsuda T., Shima E., Livio M., Anzer U., B\"orner G.,
1993, ApJ 404, 706

\bibitem[]{k58}
Kontorovich V.M., 1958, Soviet Physics JETP 6, 729, 1180

\bibitem[]{k59}
Kontorovich V.M., 1959, Soviet Physics Acoustics 5, 320

\bibitem[]{ll87}
Landau L.D., Lifshitz E.M., 1987, Fluid Mechanics, Vol. 6, Pergamon Press

\bibitem[]{m56}
McCrea W.H., 1956, ApJ 124, 461

\bibitem[]{mo83}
Meszaros P., Ostriker J.P., 1983, ApJ 273, L59

\bibitem[]{mtk99}
Molteni D., T\'oth G., Kuznetsov O.A., 1999, ApJ 516, 411

\bibitem[]{n92}
Nakayama K., 1992, MNRAS 259, 259

\bibitem[]{n93}
Nakayama K., 1993, PASJ 45, 167

\bibitem[]{n94}
Nakayama K., 1994, MNRAS 270, 871

\bibitem[]{nh94}
Nobuta K., Hanawa T., 1994, PASJ 46, 257

\bibitem[]{pom00}
Pogorelov N. V., Ohsugi Y., Matsuda, T., 2000, MNRAS 313, 198

\bibitem[]{pk83}
Protheroe R.J., Kazanas D., 1983, ApJ 265, 620

\bibitem[]{r83}
Rockwell D., 1983, AIAA Journal 21, 645

\bibitem[]{ruf94c}
Ruffert M., 1994, A\&AS~106, 505

\bibitem[]{ruf96}
Ruffert M., 1996, A\&A 311, 817

\bibitem[]{ruf94b}
Ruffert M., Arnett D., 1994, ApJ~427, 351

\bibitem[]{s81}
Scharlemann E.T., 1981, ApJ 246, L15

\bibitem[]{smabb}
Shima E., Matsuda T., Anzer U., B\"orner G., Boffin H.M.J., 1998, 
A\&A 337, 311

\bibitem[]{v83}
Vishniac E.T., 1983, ApJ 274, 152

\bibitem[]{vr89}
Vishniac E.T., Ryu D., 1989, ApJ 337, 917


\end{thebibliography}
\end{document}